\documentclass[reprint,
superscriptaddress,
 amsmath,amssymb,
 aps,
pre,
longbibliography,
]{revtex4-2}

\usepackage{graphicx}
\usepackage{dcolumn}
\usepackage{bm}
\usepackage{hyperref}

\newcommand{\dif}{\mathrm{d}}
\newcommand{\cc}{\mathbf{c}}

\newcommand{\rr}{\mathbf{r}}
\newcommand{\vv}{\mathbf{v}}
\newcommand{\uu}{\mathbf{u}}
\newcommand{\vva}{\mathbf{v}_1}
\newcommand{\vvb}{\mathbf{v}_2}
\newcommand{\vvab}{\mathbf{v}_{12}}

\newcommand{\Ssab}{\mathbf{S}_{12}}

\newcommand{\grel}{\mathbf{g}}
\newcommand{\QQ}{\mathbf{Q}}
\newcommand{\PP}{\mathsf{P}}

\newcommand{\va}{v_{1}}
\newcommand{\oa}{\omega_{1}}
\newcommand{\vb}{v_{2}}
\newcommand{\ob}{\omega_{2}}
\newcommand{\oo}{\boldsymbol{\omega}}
\newcommand{\OO}{\boldsymbol{\Omega}}
\newcommand{\ooa}{\boldsymbol{\omega}_1}
\newcommand{\oob}{\boldsymbol{\omega}_2}
\newcommand{\ww}{\mathbf{w}}
\newcommand{\s}{\widehat{\boldsymbol{\sigma}}}
\newcommand{\dt}{{d_t}}
\newcommand{\dr}{{d_r}}
\newcommand{\vth}{v_{\mathrm{th}}}
\newcommand{\oth}{\omega_{\mathrm{th}}}
\newcommand{\Dert}{\mathcal{D}_t}
\newcommand{\Tt}{T_t}
\newcommand{\Trot}{T_r}
\newcommand{\qheat}{\mathbf{q}}
\newcommand{\en}{\widetilde{\alpha}}
\newcommand{\et}{\widetilde{\beta}}
\newcommand{\een}{\alpha}
\newcommand{\eet}{\beta}
\def\a{{\alpha }}
\def\b{{\beta }}
\def\bt{\widetilde\b}
\def\at{\widetilde\a}

\DeclareMathOperator{\Tr}{Tr}

\def\bal#1\eal{\begin{align}#1\end{align}}
\newcommand\BEQ{\begin{equation}}
\newcommand{\beq}{\begin{equation}}
\newcommand\EEQ{\end{equation}}
\newcommand{\eeq}{\end{equation}}
\newcommand\beQa{\begin{eqnarray}}
\newcommand\eeQa{\end{eqnarray}}
\newcommand{\nn}{\nonumber\\}

\newcommand{\rot}{r}
\newcommand{\tr}{t}
\newcommand{\Aa}{\mathbf{A}}
\newcommand{\Ba}{\mathbf{B}}
\newcommand{\Caij}{C_{ij}}
\newcommand{\Ab}{\boldsymbol{\mathcal{A}}}
\newcommand{\Bb}{\boldsymbol{\mathcal{B}}}
\newcommand{\Cbij}{\mathcal{C}_{ij}}
\newcommand{\Eb}{\mathcal{E}}
\newcommand{\nultr}{\nu_{\lambda_\tr}}
\newcommand{\nulrot}{\nu_{\lambda_\rot}}
\newcommand{\numtr}{\nu_{\mu_\tr}}
\newcommand{\numrot}{\nu_{\mu_\rot}}
\newcommand{\nueta}{\nu_{\eta}}

\newcommand{\vom}{\bm{\Gamma}}
\newcommand{\Yt}{Y_t}
\newcommand{\Zt}{Z_t}
\newcommand{\Yr}{Y_r}
\newcommand{\Zr}{Z_r}
\newcommand{\Deltacoll}{\Delta}

\begin{document}


\title{Hydrodynamics of granular gases of inelastic and rough hard disks or spheres. I. Transport coefficients}

\author{Alberto Meg\'ias}
 \email{albertom@unex.es}
 \affiliation{Departamento de F\'isica, Universidad de Extremadura, E-06006 Badajoz, Spain}

\author{Andr\'es Santos}%
 \email{andres@unex.es}
 \affiliation{Departamento de F\'isica, Universidad de Extremadura, E-06006 Badajoz, Spain}
\affiliation{%
Instituto de Computaci\'on Cient\'ifica Avanzada (ICCAEx), Universidad de Extremadura, E-06006 Badajoz, Spain
}%

\date{\today}

\begin{abstract}
The transport coefficients for dilute granular gases of inelastic and rough  hard disks or spheres with constant  coefficients of normal ($\alpha$) and tangential ($\beta$) restitution are obtained in a unified framework as  functions of the number of translational ($d_t$) and rotational ($d_r$) degrees of freedom. The derivation is carried out by means of the  Chapman--Enskog method with a Sonine-like approximation in which, in contrast to previous approaches, the reference distribution function for angular velocities does not need to be specified. The well-known case of purely smooth $d$-dimensional particles is recovered by setting $d_t=d$ and formally taking the limit $d_r\to 0$. In addition, previous results [G.\ M.\ Kremer, A.\ Santos, and V.\ Garz\'o, Phys.\ Rev.\ E \textbf{90}, 022205 (2014)] for hard spheres are reobtained by taking $d_t=d_r=3$, while novel results for hard-disk gases are derived with the choice $d_t=2$, $d_r=1$. The singular quasismooth limit ($\beta\to -1$) and the conservative Pidduck's gas ($\alpha=\beta=1$) are also  obtained and discussed.
\end{abstract}

\maketitle



\section{Introduction}\label{sec:1}

A granular gas is essentially a system of particles that move erratically and collide inelastically. The simplest model to describe its kinetic behavior consists in a collection of inelastic hard disks (HD) or spheres (HS) with a constant  coefficient of normal restitution $\een$ (with $0\leq \een\leq 1$) \cite{C90,PL01,G03,BP04,G19}. A plausible improvement of the model is the addition of collisional friction due to surface roughness (as demanded by recent experiments \cite{YSS20}), which can be quantified via a constant  coefficient of tangential restitution $\eet$ (with $-1\leq \eet \leq 1$) \cite{JR85a}.

Certainly, this simple two-parameter model does not account for  sliding effects that can be relevant in grazing collisions \cite{FLCA94}. Models with a Coulomb friction constant \cite{W93,L99} are more realistic but less theoretically tractable outside of the quasielastic and/or quasismooth limits \cite{JZ02,GNB05,GNB05b}. Therefore, the $(\een,\eet)$ model  for granular fluids, which captures satisfactorily well the basics of collision
processes, represents an excellent compromise between simplicity and physical content  \cite{BPKZ07,GA08,MDHEH13}.

In analogy with a conventional fluid, a hydrodynamic description is also applicable and useful in the case of granular gases \cite{M93b,GS95,GZB97,SG98,D00,D01,GD02,BP03,GNB05b,SGNT06,VU09,VSG10,DB11,GS11,GSVP11a,GSVP11b,GMT13,G19}.
If the gas is made of perfectly elastic ($\alpha=1$) and either perfectly smooth ($\beta=-1$) or perfectly rough ($\beta=1$) hard particles \cite{P22}, then kinetic energy  is conserved upon collisions. Therefore, a complete set of hydrodynamic variables is defined from the densities of the conserved quantities, that is, particle density $n$ (reflecting mass conservation), flow velocity $\uu$ (due to momentum conservation), and temperature $T$ (associated with energy conservation). However, for inelastic ($\alpha\neq 1$) and/or  imperfectly rough ($|\beta|\neq 1$) hard particles, energy is no longer preserved at the collisional level. Despite that, temperature is usually included as a hydrodynamic variable \cite{G19}, except that a sink term (the so-called cooling rate) needs to be included in the energy balance equation.
Therefore, as done in Refs.\ \cite{KSG14,GSK18} for (three-dimensional) HS, in this paper we will choose $\{n,\uu,T\}$ as hydrodynamic variables. In contrast, the mean angular velocity $\OO$ is not a collisional invariant, even if $\alpha=\beta=1$, and thus it is not included as a  hydrodynamic field in our description.

To the best of our knowledge, the derivation by means of the Chapman--Enskog method of the Navier--Stokes--Fourier (NSF) hydrodynamic description (for generic constant  coefficients of restitution $\alpha$ and $\beta$) of a two-dimensional granular gas of inelastic and rough HD has not been carried out yet.  The aim of this work is to fill this gap in an inclusive way by generalizing the study to a hard-particle system with  $\dt$ and $\dr$ translational and rotational degrees of freedom, respectively, in analogy with our previous study on the energy production rates in granular mixtures \cite{MS19b,MS19}. In this way, apart from obtaining the sought results for HD gases with the choice $(\dt,\dr) = (2,1)$, the results for rough HS \cite{KSG14} are recovered by setting $(\dt,\dr) = (3,3)$. Additionally, the expressions for $d$-dimensional smooth particles ($\beta=-1$) \cite{BC01} are also reobtained by formally taking $\dr\to 0$.

Whereas the three-dimensional is perhaps the most general, verisimilar, and intuitive geometry, a two-dimensional constrained system is also found in ordinary life, like a set of marbles moving and spinning on a plane  or the pucks and strikers in the air hockey game. But the most important asset of the two-dimensional geometry  resides in its  ordinary use in experiments setups \cite{CR91,FM02,PDB03,YEHLPG11,GBM15,SP17,GBMV17,LGRAYV21,LMMRV21}. Thus, this work aims at providing testable results for the hydrodynamic transport coefficients within a general framework that encompasses the three- and two-dimensional geometries of spinning particles.

The intricacy of the $(\dt,\dr)$-generalization resides in the difficulties associated with a uniform characterization of the HS and HD vector spaces. The HS case is described by a three-dimensional Euclidean space common to both translational and angular velocities. However, to preserve the two-dimensional confinement of the HD system, angular velocities are orthogonal to the translational ones. To unify both descriptions in a common framework, we will consider the three-dimensional Euclidean space as an embedding space for the translational and angular velocity subspaces. Those subspaces coincide with the  embedding space for  HS systems, whereas  they form an orthogonal decomposition of the vector space in the HD case. Within such a description, all vector operations and relations can be written as in the HS system \cite{MS19,MS19b}. Although this mathematical description seems to be straightforward, it is rather tricky in some aspects, as will be seen.

The present paper is structured as follows. In Sec.\ \ref{sec:2}, the Boltzmann equation framework is established and the balance equations of the $\dt+2$ hydrodynamic fields, $\{n,\uu,T\}$, are derived in terms of $\dt$ and $\dr$. This mathematical description allows us to introduce in Sec.\ \ref{sec:3} the Chapman--Enskog method around the homogeneous cooling state (HCS), from which we obtain the velocity distribution function (VDF) to first order in the hydrodynamic gradients, $f^{(1)}$, under the form of four linear integral equations. To solve those equations, two successive approximations are worked out in Sec.\ \ref{sec:4}. First, a Sonine-like approximation is assumed without prejudicing the form of the zeroth-order HCS VDF $f^{(0)}$; this allows us to express the NSF transport coefficients in terms of velocity cumulants  and collision integrals of $f^{(0)}$. As a second step,  the unknown function $f^{(0)}$ is approximated by  a Maxwellian distribution for the translational velocities times a generic marginal distribution for the angular velocities, what allows us to derive explicit expressions for the transport coefficients (see Table \ref{table0} below).  The results are illustrated in Sec.\ \ref{sec:5} for both spheres and disks, including some interesting limiting situations. Finally, concluding remarks and main results are summed up in Sec.\ \ref{sec:6}.

\section{Granular gas of inelastic and rough hard particles}\label{sec:2}
\subsection{Boltzmann equation}
We consider a HD or HS granular gas made of identical particles of diameter $\sigma$, mass $m$, and moment of inertia $I=\kappa m\sigma^2/4$. The reduced moment of inertia takes the values $\kappa = \frac{1}{2}$ for uniform disks and $\kappa = \frac{2}{5}$ for uniform spheres; its maximum value is $\kappa_{\max}=1$ (HD) and $\kappa_{\max}=\frac{2}{3}$ (HS). The translational and angular velocities of a particle will be denoted by $\vv$ and $\oo$, respectively.
Whenever convenient, we will use the short-hand notations $\vom\equiv\{\vv,\oo\}$ and $\int\dif\vom\equiv \int\dif\vv\int\dif\oo$ for simplicity.

Particle-particle collisions are characterized by constant coefficients of normal ($\alpha$) and tangential ($\beta$) restitution (see Appendix \ref{appA} for a summary of the collision rules).
As said in Sec.\ \ref{sec:1},  vector relations within our generalized description belongs to an embedding space, namely the three-dimensional Euclidean space $\mathfrak{E}$. Therefore, the collision rules in Appendix \ref{appA} are presented in the three-dimensional framework.

We will carry out a kinetic-theory description of a dilute granular gas, in the sense that the one-body VDF will be enough to characterize the system. This approach is complemented with the assumption of molecular chaos or \emph{Stosszahlansatz}. The analytical treatment is then based on the Boltzmann equation in the absence of external forces, which reads
\BEQ\label{eq:8}
    \partial_t f+\vv\cdot\nabla f= J_{\vom}[f,f],
\EEQ
where  $f=f(\rr,\vom;t)$ is the VDF at time $t$ and $J_{\vom}$ is the Boltzmann bilinear collision operator:
\bal\label{eq:9}
    J_{\vom_1}[f,f]=&\sigma^{\dt-1}\int \dif\vom_2\int_{+} \dif \s \medspace (\vvab\cdot \s)\nn
    &\times \left[\frac{f_1^{ \prime\prime}f_2^{\prime\prime}}{\alpha^2 |\beta|^{2\dr/\dt}}-f_1 f_2 \right].
\eal
Here, $\vvab\equiv \vva-\vvb$ is the relative translational velocity, $\s=(\mathbf{r}_2-\mathbf{r}_1)/|\mathbf{r}_2-\mathbf{r}_1|$ is the  intercenter unit vector at contact, the subscript $+$ in the integral over $\s$ designates  the constraint $\s\cdot\vvab >0$, and $f_{1,2} = f(\vom_{1,2})$ and $f_{1,2}^{\prime\prime} = f(\vom_{1,2}^{\prime\prime})$, the double primes denoting precollisional quantities giving rise to unprimed quantities as postcollisional values. Moreover, use has been made of the Jacobian given by \eqref{Jacob}.

\subsection{Hydrodynamic balance equations}\label{subsec:2.2}
From a macroscopic point of view, the flow of a low-density granular gas can be fully described by the knowledge of the following hydrodynamic fields: particle number density $n(\rr,t)$, hydrodynamic flow velocity $\uu(\rr,t)$, and  total granular temperature $T(\rr,t)$. They are given by
\begin{subequations}
\BEQ\label{eq:11}
    n(\rr,t) = \int \dif \vom \medspace f(\rr,\vom;t),
\EEQ
\beq
\uu(\rr,t)=\langle \vv\rangle,
\eeq
\BEQ\label{eq:14}
    T(\rr,t) = \frac{\dt \Tt(\rr,t)+ \dr\Trot(\rr,t)}{\dt+\dr},
\EEQ
\end{subequations}
where
\BEQ\label{eq:12}
    \Tt(\rr,t) = \frac{m}{\dt}\langle V^2\rangle, \quad \Trot(\rr,t) = \frac{I}{\dr}\langle \oo^2\rangle,
\EEQ
$\mathbf{V}= \vv-\uu$ being the peculiar velocity. The angular brackets denote averages defined generically as
\BEQ\label{eq:13}
    \langle \psi \rangle = \frac{1}{n(\rr,t)}\int\dif\vom \medspace \psi(\rr,\vom;t) f(\rr,\vom;t).
\EEQ
Note that the rotational temperature $\Trot$ is not defined with respect to the mean angular velocity $\OO = \langle \oo \rangle$ because the latter is not a conserved quantity \cite{KSG14}.

Given a quantity $\psi(\rr,\vom;t)$, its associated transfer equation can be obtained by multiplying both sides of the Boltzmann equation, Eq.\ \eqref{eq:8}, by $\psi$  and integrating over translational and angular velocities. The result is
\BEQ\label{eq:15}
    \partial_t(n\langle \psi\rangle) +\nabla\cdot (n\langle \vv\psi\rangle) -n \langle (\partial_t+\vv\cdot\nabla) \psi\rangle =  \mathcal{J}[\psi|f,f],
\EEQ
where $\mathcal{J}[\psi|f,f]$ is the collisional production term of the quantity $\psi$,  given by
\bal\label{eq:16}
    \mathcal{J}[\psi|f,f] =& \int\dif\vom_1 \medspace \psi(\rr_1,\vom_1;t)J_{\vom_1}[f,f] \nn
    =& \frac{\sigma^{\dt-1}}{2} \int\dif\vom_1\int\dif\vom_2\int_{+}\dif\s\,(\s\cdot\vvab) \nn
    &\times \Deltacoll(\psi_{1}+\psi_{2}) f_1 f_2.
\eal
Here, the operator $\Deltacoll$ acting on a generic quantity $\psi$  yields the difference between the postcollisional and precollisional values of $\psi$, i.e., $\Deltacoll \psi(\vom) \equiv \psi(\vom')-\psi(\vom)$.

The balance equations for mass, momentum, and energy are obtained from Eq.\ \eqref{eq:15} by choosing $\psi=1$, $\psi=m\vv$, $\psi = \frac{1}{2}m V^2$, $\psi = \frac{1}{2}I \omega^2$, and $\psi =\frac{1}{2}m V^2+ \frac{1}{2}I \omega^2$. This yields, respectively,
\begin{subequations}
       \BEQ\label{eq:17}
        \Dert n +n \nabla\cdot\uu = 0,
        \EEQ
        \BEQ\label{eq:18}
        \Dert \uu +\rho^{-1}\nabla\cdot\PP = 0,
        \eeq
        \BEQ\label{eq:19}
        \Dert \Tt +\frac{2}{n\dt}(\nabla \cdot \mathbf{q}_\tr + \PP: \nabla \uu) +\Tt\zeta_\tr = 0,
    \EEQ
     \BEQ\label{eq:21}
        \Dert \Trot +\frac{2}{n\dr}\nabla \cdot \mathbf{q}_\rot +\Trot\zeta_\rot= 0,
    \EEQ
       \BEQ\label{eq:23}
        \Dert T +\frac{2}{n(\dt+\dr)}(\nabla \cdot \mathbf{q} +\PP : \nabla \uu) +T\zeta= 0.
    \EEQ
    \end{subequations}
In these equations, $\Dert\equiv \partial_t+\uu\cdot\nabla$ is the material time derivative and $\rho\equiv mn$ is the mass density. Moreover, $\mathsf{P}$ is the pressure tensor,
$\qheat_\tr$ ($\qheat_\rot$) is the translational (rotational) contribution to the total heat flux $\qheat$, $\zeta_\tr$ ($\zeta_\rot$) is the translational (rotational) energy production rate, and $\zeta$ is the cooling rate. These quantities are defined as
    \begin{subequations}
    \label{eq:25-24}
    \BEQ\label{eq:25}
    P_{ij} = \rho\langle V_iV_j\rangle,\quad  p = \frac{1}{\dt}\Tr\mathsf{P} = n\Tt,
\EEQ
    \BEQ\label{eq:26}
    \qheat_\tr = \frac{\rho}{2}\langle V^2\mathbf{V}\rangle, \quad \qheat_\rot = \frac{In}{2}\langle \omega^2 \mathbf{V}\rangle ,\quad \qheat = \qheat_\tr+\qheat_\rot,
\EEQ
\BEQ\label{eq:20}
        \zeta_\tr = - \frac{m}{\dt n\Tt} \mathcal{J}[v^2|f,f],\quad \zeta_\rot = - \frac{I}{\dr n\Trot} \mathcal{J}[\omega^2|f,f],
    \EEQ
        \beq\label{eq:24}
        \zeta = \frac{\dt\zeta_\tr\Tt+\dr\zeta_\rot\Trot}{(\dt+\dr)T}
       .
    \eeq
\end{subequations}
In Eq.\ \eqref{eq:20}, the collisional rates of change $\mathcal{J}[v^2|f,f]$ and $\mathcal{J}[\omega^2|f,f]$ are obtained from Eq.\ \eqref{eq:16} by setting $\psi=v^2$ and $\psi=\omega^2$, respectively.

\subsection{Homogeneous cooling state}
Before analyzing inhomogeneous states  in terms of the transport coefficients at the NSF order in Secs.\ \ref{sec:3}--\ref{sec:5},
let us consider the HCS, henceforth represented by the superscript $(0)$. In that case ($\nabla\to 0$), Eqs.\ \eqref{eq:17} and \eqref{eq:18} yield
$n=\text{const}$ and $\uu=\text{const}$, while
Eqs.\ \eqref{eq:19}--\eqref{eq:23} become
\begin{subequations}
\label{eq:n19-n23}
\BEQ
\label{eq:n19}
         \dot{T}_\tr^{(0)} +{T}_\tr^{(0)}\zeta_\tr^{(0)} = 0,
    \EEQ
     \BEQ\label{eq:n21}
        \dot{T}_\rot^{(0)} +{T}_\rot^{(0)}\zeta_\rot^{(0)}= 0,
    \EEQ
       \BEQ\label{eq:n23}
        \dot{T} +T\zeta^{(0)}= 0.
    \EEQ
\end{subequations}
Note that we have not attached a superscript $(0)$ to the global temperature $T$ because of its status as a hydrodynamic variable. The Boltzmann equation, Eq.\ \eqref{eq:8},  reduces in the HCS to
\BEQ\label{eq:n8}
    \partial_t f^{(0)}(\vom;t)=J_{\vom}[f^{(0)},f^{(0)}].
\EEQ
Since in the HCS all the time-dependence of $f^{(0)}$ occurs through a dependence on $T$, we can write \cite{G19,KSG14}
\beq
\label{eq:n8b}
\partial_t f^{(0)}=\dot{T}\frac{\partial f^{(0)}}{\partial T}=\frac{\zeta^{(0)}}{2}\left(\frac{\partial}{\partial\mathbf{V}}\cdot\mathbf{V}+\frac{\partial}{\partial\oo}\cdot\oo\right)f^{(0)}.
\eeq

The rotational-to-translational, translational-to-total, and rotational-to-total temperature ratios  are defined as
\begin{subequations}
\beq
\theta\equiv \frac{T_\rot^{(0)}}{T_\tr^{(0)}},
\eeq
\beq
\label{tau_t}
\tau_\tr\equiv\frac{T_\tr^{(0)}}{T}=\frac{\dt+\dr}{\dt+\dr\theta},\quad \tau_\rot\equiv\frac{T_\rot^{(0)}}{T}=\frac{\dt+\dr}{\dt/\theta+\dr}.
\eeq
\end{subequations}
Those temperature ratios are stationary in the HCS, so that Eqs.\ \eqref{eq:n19-n23} imply that $\zeta_\tr^{(0)}=\zeta_\rot^{(0)}=\zeta^{(0)}$.

The exact solution to Eq.\ \eqref{eq:n8} is not known, but good estimates for the production rates $\zeta_\tr^{(0)}$, $\zeta_\rot^{(0)}$, and $\zeta^{(0)}$ can be obtained by assuming the simple trial function
\beq
f^{(0)}(\vom)\to  n \vth^{-\dt}\oth^{-\dr}\pi^{-\dt/2}e^{-c^2}\varphi_r(\ww),
\label{3.1}
\eeq
where
\beq
\vth = \sqrt{\frac{2T_\tr^{(0)}}{m}},\quad \oth = \sqrt{\frac{2T_\rot^{(0)}}{I}}
\eeq
are the translational and rotational thermal velocities, and
\BEQ
\label{cc_ww}
    \cc = \frac{\mathbf{V}}{\vth}, \quad \ww = \frac{\oo}{\oth}
\EEQ
are the scaled translational and angular velocities. Note that, while a Maxwellian translational distribution has been assumed, the (isotropic) marginal rotational distribution $\varphi_r(\ww)$ does not need to be specified.
Within this approximation, the results are \cite{MS19,MS19b}
\begin{subequations}
\label{6abc}
  \bal
   \zeta_\tr^{(0)}=&\frac{\nu}{\dt}\Big\{1-\alpha^2+\frac{2\dr\kappa(1+\beta)}{\dt(1+\kappa)^2 }\Big[1-\theta
   \nn
&+\frac{\kappa(1-\beta)}{2}\left(1
  +\frac{\theta}{\kappa}\right)\Big]\Big\},
\eal
\beq
\zeta_\rot^{(0)}=\frac{2\nu}{\dt}\frac{\kappa(1+\beta)}{(1+\kappa)^2}\left[1-\frac{1}{\theta}+
\frac{1-\beta}{2}\left(\frac{1}{\theta}+\frac{1}{\kappa}\right)\right],
\eeq
\BEQ
\label{eq:38}
    \zeta^{(0)} = \frac{\nu}{\dt+\dr\theta}\left[1-\een^2+\frac{\dr}{\dt}\frac{1-\eet^2}{1+\kappa}\left({\kappa}+{\theta}\right)\right],
\EEQ
\end{subequations}
where  $\nu$ is the collision frequency defined as
\BEQ\label{eq:40}
    \nu = K  n \sigma^{\dt-1}\vth, \quad K\equiv \frac{\sqrt{2}\pi^{\frac{\dt-1}{2}}}{\Gamma\left(\dt/2\right)}.
\EEQ
Note that $\nu=\frac{\dt+2}{4}\nu_0$, where $\nu_0$ is the collision frequency associated with the shear viscosity of a molecular gas \cite{GSM07}.
Insertion of Eqs.\ \eqref{6abc} into the condition $\zeta_\tr^{(0)}=\zeta_\rot^{(0)}$ yields the quadratic equation $\theta-1-(\dt/\dr)(1/\theta-1)=2h$, where
\beq
h\equiv\frac{\dt(1+\kappa)^2}{2\dr\kappa(1+\beta)^2}\left[1-\alpha^2-\frac{1-\frac{\dr}{\dt}\kappa}{1+\kappa}(1-\beta^2) \right],
\eeq
whose physical solution is
\beq
\label{theta}
\theta=\sqrt{\left[h- \frac{1}{2}\left(\frac{\dt}{\dr}-1\right)\right]^2+\frac{\dt}{\dr}}+h-\frac{1}{2}\left(\frac{\dt}{\dr}-1\right).
\eeq

\section{Chapman--Enskog method}
\label{sec:3}

The main goal of this paper is to obtain the NSF constitutive equations with explicit expressions for the associated transport coefficients. As usual, this will be done by assuming that the VDF depends on space and time only through the \emph{slow} hydrodynamic fields introduced before ($n$, $\uu$, and $T$) and applying  the Chapman--Enskog  expansion method \cite{BP04,G19}.

\subsection{General scheme}

The Chapman--Enskog method consists essentially in introducing multi-scale space-time derivatives and a perturbation expansion of the VDF in powers of the gradients of the hydrodynamic fields, namely
\begin{subequations}
\beq
 \nabla \rightarrow \epsilon \nabla,\quad    f = f^{(0)} + \epsilon f^{(1)} + \epsilon^2 f^{(2)}+\cdots, \label{eq:29a}
\eeq
\beq
    \Dert = \Dert^{(0)}+\epsilon\Dert^{(1)}+\epsilon^2\Dert^{(2)}+\cdots, \label{eq:29b}
\eeq
\end{subequations}
where $\epsilon$ is a bookkeeping parameter. Thus, the Boltzmann equation, Eq.\ \eqref{eq:8}, decouples into a hierarchy of equations of orders $k=0,1,2,\ldots$. The zeroth- and first-order equations are
\begin{subequations}
\beq
    \Dert^{(0)} f^{(0)} = J_{\vom}[f^{(0)},f^{(0)}], \label{eq:35a}
\eeq
\beq
    \left(\Dert^{(0)}+\mathcal{L}\right)f^{(1)} = - \left(\Dert^{(1)}+\mathbf{V}\cdot\nabla\right) f^{(0)}. \label{eq:35b}
\eeq
\end{subequations}
In Eq.\ \eqref{eq:35b}, the linear collision operator $\mathcal{L}$ is defined as
\bal\label{eq:33}
    \mathcal{L} \Phi(\vom_1)=-J_{\vom_1}[\Phi,f^{(0)}]-J_{\vom_1}[f^{(0)},\Phi].
\eal
Comparison between Eqs.\ \eqref{eq:n8} and \eqref{eq:35a} shows that the zeroth-order VDF $f^{(0)}$ is the local version of the HCS VDF. This will be further confirmed below.

Substituting Eq.\ \eqref{eq:29a} into Eqs.\ \eqref{eq:25-24}, one obtains
\begin{subequations}
\beq
    P_{ij} = p^{(0)}\delta_{ij}+\epsilon P_{ij}^{(1)}+\epsilon^2 P_{ij}^{(2)}+\cdots,\label{eq:30a}
\eeq
\beq
    \qheat = \epsilon \qheat^{(1)}+\epsilon^2 \qheat^{(2)}+\cdots,\label{eq:30b}
\eeq
\beq
   \zeta = \zeta^{(0)}+\epsilon \zeta^{(1)}+\epsilon^2 \zeta^{(2)}+\cdots.\label{eq:30c}
\eeq
\end{subequations}
Here, $p^{(0)}=n\tau_t T$, $\tau_\tr$ being defined by Eq.\ \eqref{tau_t} and
\begin{subequations}
\label{eq:32}
\beq
    \zeta^{(1)} =\frac{\dt \tau_\tr \zeta_\tr^{(1)}+\dr \tau_\rot \zeta_\rot^{(1)}}{\dt+\dr},
\eeq
\beq
\zeta_\tr^{(1)}=-\frac{m}{\dt n\tau_\tr T}\Lambda[V^2|f^{(1)}],\quad \zeta_\rot^{(1)}=-\frac{I}{\dr n\tau_\rot T}\Lambda[\omega^2|f^{(1)}],
\eeq
\end{subequations}
where, in general,
\bal
\label{Lambda}
\Lambda[\psi|\Phi]\equiv &\int\dif\vom_1 \, \psi(\vom_1)\mathcal{L}\Phi(\vom_1)\nn
=&-\frac{\sigma^{\dt-1}}{2}\int\dif\vom_1\int\dif\vom_2\int_+\dif\s\, (\s\cdot\vvab)\nn
&\times \Deltacoll\left(\psi_1+\psi_2\right)\left(f_1^{(0)}\Phi_2+\Phi_1f_2^{(0)}\right).
\eal
Note that, within the approximation described by Eq.\ \eqref{3.1}, $\theta$ and $\zeta^{(0)}$ are given by Eqs.\ \eqref{theta} and \eqref{eq:38}, respectively.

Furthermore, the action of the operator $\Dert^{(k)}$ on a generic function $\psi(n,\uu,T)$ of the hydrodynamic fields is
\BEQ\label{eq:36}
    \Dert^{(k)}\psi = \frac{\partial \psi}{\partial n}\Dert^{(k)}n+\frac{\partial \psi}{\partial \uu}\cdot \Dert^{(k)}\uu +\frac{\partial \psi}{\partial T}\Dert^{(k)}T,
\EEQ
where $\Dert^{(k)}n$, $\Dert^{(k)}\uu$, and $\Dert^{(k)}T$ are obtained from the balance equations, Eqs.\ \eqref{eq:17}, \eqref{eq:18}, and \eqref{eq:23}. In particular,
\begin{subequations}
\label{eq:37a-c}
\beq
\label{eq:37a}
    \Dert^{(0)}n = 0, \quad \Dert^{(0)}\uu = 0, \quad \Dert^{(0)}T =-T\zeta^{(0)},
\eeq
\beq
    \Dert^{(1)}n = -n\nabla\cdot\uu, \quad \Dert^{(1)}\uu=-\frac{\tau_\tr}{\rho}\nabla (nT),
    \label{eq:37b}
\eeq
\beq
 \Dert^{(1)}T=-\frac{2\tau_\tr}{\dt+\dr} T \nabla\cdot\uu-T\zeta^{(1)}. \label{eq:37c}
\eeq
\end{subequations}
Equation \eqref{eq:37a} implies that  $\Dert^{(0)}f^{(0)} = -\zeta^{(0)}T\partial_T f^{(0)}$, in agreement with Eq.\ \eqref{eq:n8b}. This confirms that $f^{(0)}$ is the local version of the HCS VDF.

\subsection{First-order distribution}

By following the same steps as in Sec.\ IVB of Ref.\ \cite{KSG14}, it is possible to express the solution to Eq.\ \eqref{eq:35b} as
\bal\label{eq:44}
    f^{(1)} =& \Ab\cdot\nabla \ln T+ \Bb\cdot\nabla \ln n+ \Cbij \nabla_j u_i+\Eb \nabla\cdot\uu,
\eal
where the  functions $\Ab$, $\Bb$, $\Cbij$, and $\Eb$ obey the following set of linear integral equations:
\begin{subequations}
\label{eq:48a-d}
\beq
    \left(-\frac{\zeta^{(0)}}{2}-\zeta^{(0)}T\partial_T+\mathcal{L}\right)\Ab = \Aa, \label{eq:48a}
\eeq
\beq
    \left(-\zeta^{(0)}T\partial_T+\mathcal{L}\right)\Bb -\zeta^{(0)}\Ab= \Ba , \label{eq:48b}
\eeq
\beq
    \left(-\zeta^{(0)}T\partial_T+\mathcal{L} \right)\Cbij = \Caij, \label{eq:48c}
\eeq
\beq
    \left(-\zeta^{(0)}T\partial_T+\mathcal{L}\right)\Eb + \xi T\partial_T f^{(0)} = E. \label{eq:48d}
\eeq
\end{subequations}
Here, the functions in the inhomogeneous terms are defined by the relation $-(\Dert^{(1)}+\mathbf{V}\cdot\nabla) f^{(0)} =\Aa\cdot \nabla \ln T + \Ba\cdot\nabla \ln n+\Caij \nabla_j u_i +E \nabla\cdot\uu + \zeta^{(1)}T \partial_T f^{(0)}$. They are  given by
\begin{subequations}
\label{eq:42a-d}
\beq
        \Aa = -\frac{\vth}{2}\left(\partial_\cc-\cc\partial_\cc\cdot\cc-\cc\partial_{\ww}\cdot\ww\right)f^{(0)},\label{eq:42a}
\eeq
\beq
        \Ba = -\frac{\vth}{2}\left(2\cc+\partial_\cc \right)f^{(0)}, \label{eq:42b}
\eeq
\beq
        \Caij = -\left(\frac{1}{\dt}\delta_{ij}\cc\cdot\partial_\cc-c_j\partial_{c_i}\right)f^{(0)}, \label{eq:42c}
\eeq
\beq
        E = -\frac{\dr\tau_\tr\tau_\rot}{\dt+\dr}\left(\frac{\partial_{\ww}\cdot\ww}{\dr\tau_\rot}-\frac{\dt+\cc\cdot\partial_\cc}{\dt\tau_\tr}\right)f^{(0)}, \label{eq:42d}
\eeq
\end{subequations}
where $\delta_{ij}$ is the identity tensor in the translational velocity Euclidean subspace, the scaled velocities $\cc$ and $\ww$ are defined by Eq.\ \eqref{cc_ww}, and use has been made of
the general property $T\partial_T f^{(0)}=-\frac{1}{2}\left(\partial_{\mathbf{V}}\cdot\mathbf{V}+\partial_{\oo}\cdot\oo\right)f^{(0)}$ [see the second equality in Eq.\ \eqref{eq:n8b}].
In Eq.\ \eqref{eq:48d}, $\xi$ is the velocity-divergence transport coefficient in the constitutive equation
\beq
\label{eq:45c}
\zeta^{(1)} = -\xi\nabla\cdot\uu,
\eeq
which is given by
\begin{subequations}
\label{eq:45}
\BEQ
\label{eq:45a}
 \xi = \frac{\dt\tau_\tr\xi_\tr+\dr\tau_\rot\xi_\rot}{\dt+\dr},
\EEQ
\beq
\xi_\tr=-\frac{m}{\dt n\tau_\tr T}\Lambda[V^2|\Eb],\quad \xi_\rot=-\frac{I}{\dr n\tau_\rot T}\Lambda[\omega^2|\Eb],
\label{eq:45b}
\eeq
\end{subequations}

Note that $\Caij$ is a traceless tensor. However, in general, it is not symmetric.  In the HS case, due to isotropy, the local version of the HCS function  $f^{(0)}$ is a function of $V^2$, $\omega^2$, and $\vartheta \equiv (\mathbf{V}\cdot\oo)^2$ \cite{SKS11,VSK14,VSK14b}. This implies \cite{KSG14} $\Caij -C_{ji} = 2\left({\partial f^{(0)}}/{\partial \vartheta}\right)(\mathbf{V}\cdot\oo)(V_j\omega_i-V_i\omega_j)$.
However,  the vectors $\mathbf{V}$ and $\oo$ are mutually orthogonal in the HD case and hence the tensor $\Caij$ is symmetric in the two-dimensional geometry.

\subsection{Navier--Stokes--Fourier transport coefficients}
\label{sec:3C}

The formal derivation from Eq.\ \eqref{eq:44} of the constitutive equations for the pressure tensor and the heat flux follows the same steps as in Sec.\ V of Ref.\ \cite{KSG14}, except that special care must be exerted to redo those steps keeping $\dt$ and $\dr$ generic. For the sake of conciseness, we skip some of the technical details.

The first-order  pressure tensor and  heat flux can be expressed as
\begin{subequations}
\label{eq:52-53}
\BEQ
\label{eq:52}
    P_{ij}^{(1)} = -\eta\left(\nabla_i u_j+\nabla_j u_i -\frac{2}{\dt}\delta_{ij}\nabla\cdot\uu\right)-\eta_b \delta_{ij}\nabla\cdot\uu,
\EEQ
\BEQ\label{eq:53}
    \qheat^{(1)} = -\lambda\nabla T-\mu \nabla n,
\EEQ
\end{subequations}
where $\eta$ is the sear viscosity, $\eta_b$ is the bulk viscosity, $\lambda$ is the thermal conductivity, and $\mu$ is a Dufour-like cofficient \cite{BC01,GTSH12,KSG14,GSK18,BP03,G19}.
Since $\qheat^{(1)}$ has a translational and a rotational contribution [see Eq.\ \eqref{eq:26}] so do $\lambda$ and $\mu$:
\beq
       \lambda =\tau_\tr\lambda_\tr+ \tau_\rot\lambda_\rot, \quad    \mu = \mu_\tr+\mu_\rot. \label{eq:57}
\eeq

The transport coefficients can be expressed in terms of the solutions to Eqs.\ \eqref{eq:48a-d} as
\begin{subequations}
\label{eq:73-65}
\bal
     \eta =&-\frac{m}{(\dt+2)(\dt-1)}\int\dif\vom \, \left(V_iV_j-\frac{1}{\dt}\delta_{ij}V^2 \right)\Cbij\nn
     =& \frac{n\tau_\tr T}{\nueta-\frac{1}{2}\zeta^{(0)}},\label{eq:73}
\eal
\beq
     \eta_b =-\frac{m}{\dt}\int\dif\vom\, V^2 \Eb=\frac{\tau_\tr\tau_\rot nT}{\zeta^{(0)}} \frac{2\dr}{\dt+\dr}\left(\xi_\tr-\xi_\rot-\frac{2}{\dt}\right), \label{eq:74}
\eeq
\beq
    \lambda_\tr =-\frac{m}{2\dt \tau_\tr T}\int\dif\vom\, V^2\mathbf{V}\cdot\Ab= \frac{\dt+2}{2}\frac{n\tau_\tr T}{m} \frac{1+2a_{20}^{(0)}}{\nultr-2\zeta^{(0)}}, \label{eq:62}
\eeq
\beq
    \lambda_\rot=  -\frac{I}{2\dt \tau_\rot T}\int\dif\vom\, \omega^2\mathbf{V}\cdot\Ab= \frac{\dr}{2}\frac{n\tau_\tr T}{m} \frac{1+2a_{11}^{(0)}}{\nulrot-2\zeta^{(0)}},\label{eq:63}
\eeq
\beq
    \mu_\tr = -\frac{m}{2\dt n}\int\dif\vom\, V^2\mathbf{V}\cdot\Bb
    =\frac{\tau_\tr T}{n}\frac{\lambda_\tr\zeta^{(0)}+\frac{\dt+2}{2}\frac{n\tau_\tr T }{m}a_{20}^{(0)}}{\numtr-\frac{3}{2}\zeta^{(0)}},\label{eq:64}
\eeq
\beq
    \mu_\rot = -\frac{I}{2\dt n}\int\dif\vom\, \omega^2\mathbf{V}\cdot\Bb
    =\frac{\tau_\rot T}{n}\frac{\lambda_\rot\zeta^{(0)}+\frac{\dr}{2}\frac{n\tau_\tr T }{m}a_{11}^{(0)}}{\numrot-\frac{3}{2}\zeta^{(0)}}, \label{eq:65}
\eeq
\end{subequations}
where we have introduced the HCS cumulants
\begin{subequations}
\label{eq:66-67}
\beq
    a_{20}^{(0)} = \frac{m^2}{\dt(\dt+2)\tau_\tr^2T^2}\langle V^4\rangle^{(0)}-1, \label{eq:66}
\eeq
\beq
    a_{11}^{(0)} = \frac{mI}{\dr\dt\tau_\tr \tau_\rot T^2}\langle V^2\omega^2\rangle^{(0)}-1, \label{eq:67}
\eeq
\end{subequations}
and the collision frequencies
\begin{subequations}
\label{eq:69-72}
\beq
\nueta=\frac{\Lambda[V_iV_j-\frac{1}{\dt}V^2\delta_{ij}|\mathcal{C}_{ij}]}{\int\dif\vom\,\left(V_iV_j-\frac{1}{\dt}V^2\delta_{ij}\right)\mathcal{C}_{ij}},
\eeq
\beq
    \nultr = \frac{\Lambda[V^2{V}_i|\mathcal{A}_i]}{\int\dif\vom\, V^2\mathbf{V}\cdot\Ab}, \quad
   \nulrot = \frac{\Lambda[\omega^2{V}_i|\mathcal{A}_i]}{\int\dif\vom\, \omega^2\mathbf{V}\cdot\Ab}, \label{eq:69}
\eeq
\beq
\numtr = \frac{\Lambda[ V^2{V}_i|\mathcal{B}_i]}{\int\dif\vom\, V^2\mathbf{V}\cdot\Bb}, \quad
\numrot = \frac{\Lambda[\omega^2{V}_i|\mathcal{B}_i]}{\int\dif\vom\, \omega^2\mathbf{V}\cdot\Bb}. \label{eq:72}
\eeq
\end{subequations}

It is interesting to remark that the rotational-to-translational temperature ratio is affected by the presence of $\nabla\cdot\uu$. Taking the trace in both sides of Eq.\ \eqref{eq:52}, we get, to first order, $\Tt=\tau_\tr T-(\eta_b/n)\nabla\cdot\uu$. Since Eq.\ \eqref{eq:14} must hold to any order, this implies $\Trot=\tau_\rot T+(\dt/\dr)(\eta_b/n)\nabla\cdot\uu$. As a consequence,
\BEQ\label{eq:75}
    \frac{\Trot}{\Tt} = \theta+ \frac{\tau_\rot}{\tau_\tr} \frac{2}{\zeta^{(0)}} \left( \xi_\tr-\xi_\rot-\frac{2}{\dt} \right)\nabla\cdot\uu.
\EEQ

\section{Explicit Expressions for the Transport Coefficients}
\label{sec:4}

All the expressions in Sec.\ \ref{sec:3} are formally exact within the Chapman--Enskog scheme but they are not explicit since neither the zeroth-order VDF $f^{(0)}$ nor the solutions to the linear integral equations \eqref{eq:48a-d} are known exactly.

By symmetry arguments,  $\Ab$ and $\Bb$ can be expressed, in the HS case, as linear combinations of the vectors $\mathbf{V}$, $(\mathbf{V}\cdot\oo)\oo$, and $\mathbf{V}\times\oo$, while $\Cbij$ is a linear combination of the dyadic products of those three vectors.
However, $\mathbf{V}\perp\oo$ in a HD system,  and thus $\Ab$ and $\Bb$ are vector functions  residing in the two-dimensional subspace  $\mathfrak{V}$ of translational velocities, so that they can be expressed as linear combinations of the mutually orthogonal vectors $\mathbf{V}$ and $\mathbf{V}\times \oo$ (which form an orthogonal basis of $\mathfrak{V}$), where the latter vector product is done in the embedding space $\mathfrak{E}=\mathfrak{V}\oplus \mathfrak{W}$, $\mathfrak{W}$ being the one-dimensional subspace where angular velocities live. Then, in the case of disks, $\Cbij$ is a linear combination of the dyadic products of the two vectors $\mathbf{V}$ and $\mathbf{V}\times \oo$ only.

\subsection{Sonine-like approximation for $\mathcal{A}_i$, $\mathcal{B}_i$   $\Cbij$, and $\Eb$}\label{sec:IV.A}

To get explicit expressions for the NSF transport coefficients we need to resort to approximations. We will proceed in two steps. First,
the structure of Eqs.\ \eqref{eq:42a-d} suggests to propose the following approximate forms for the solutions of Eqs.\ \eqref{eq:48a-d}:
\begin{subequations}
\label{Sonine}
\beq
        \Ab\to -\frac{\vth}{2\nu}\left[\gamma_{A_t}\left(\partial_\cc-\cc\partial_\cc\cdot\cc\right)-\gamma_{A_r}\cc\partial_{\ww}\cdot\ww\right]f^{(0)},
\eeq
\beq
        \Bb\to -\frac{\vth}{2\nu}\left[\gamma_{B_t}\left(\partial_\cc-\cc\partial_\cc\cdot\cc\right)-\gamma_{B_r}\cc\partial_{\ww}\cdot\ww\right]f^{(0)},
\eeq
\beq
        \Cbij \to -\frac{\gamma_C}{2\nu}\left(\frac{1}{\dt}\delta_{ij}\cc\cdot\partial_\cc-c_j\partial_{c_i}\right)f^{(0)},
\eeq
\beq
        \Eb\to -\frac{\gamma_E\dr\tau_\tr\tau_\rot}{2\nu}\left(\frac{\partial_{\ww}\cdot\ww}{\dr\tau_\rot}-\frac{\dt+\cc\cdot\partial_\cc}{\dt\tau_\tr}\right)f^{(0)},
\eeq
\end{subequations}
where $\nu$ is defined by Eq.\ \eqref{eq:40} and the $\gamma$ coefficients remain to be determined. In the case of conservative collisions ($\alpha=|\beta|=1$), $f^{(0)}$ is the Maxwellian equilibrium distribution and then Eqs.\ \eqref{Sonine} define the simplest Sonine approximation \cite{CC70,K10a}. Therefore, Eqs.\ \eqref{Sonine} will be referred to as Sonine-like approximation.

Inserting Eqs.\ \eqref{Sonine} into the first equalities in Eqs.\ \eqref{eq:73-65}, one can relate the transport coefficients to the $\gamma$ coefficients as follows:
\begin{subequations}
\label{4.2a-e}
\beq
    \eta^*\equiv\frac{\eta}{\eta_0} = \frac{2\gamma_C}{\dt+2}, \quad \eta_b^*\equiv\frac{\eta_b}{\eta_0} = \frac{4\dr\tau_\rot  \gamma_E}{\dt(\dt+2)},
\eeq
\beq
    \lambda_\tr^*\equiv\frac{\lambda_\tr}{\lambda_0} = \frac{4(\dt-1)}{\dt(\dt+2)}[1+2a_{20}^{(0)}]\gamma_{A_\tr},
\eeq
\beq
    \lambda_\rot^*\equiv\frac{\lambda_\rot}{\lambda_0} = \frac{4(\dt-1)\dr}{\dt(\dt+2)^2}\left[\gamma_{A_\rot}+\left(\gamma_{A_\rot}+{\gamma_{A_\tr}}\right)a_{11}^{(0)}\right],
\eeq
\beq
    \mu_\tr^*\equiv\frac{n\mu_\tr}{\lambda_0 T} = \frac{4(\dt-1)\tau_\tr}{\dt(\dt+2)}[1+2a_{20}^{(0)}]\gamma_{B_\tr},
\eeq
\beq
\mu_\rot^*\equiv\frac{n\mu_\rot}{\lambda_0 T} = \frac{4(\dt-1)\dr\tau_\rot}{\dt(\dt+2)^2}\left[\gamma_{B_\rot}+\left(\gamma_{B_\rot}+{\gamma_{B_\tr}}\right)a_{11}^{(0)}\right],
\eeq
\end{subequations}
where
\BEQ
\label{eta0}
    \eta_0 = \frac{\dt+2}{4}\frac{n\tau_\tr T}{\nu}, \quad \lambda_0 = \frac{\dt(\dt+2)}{2(\dt-1)}\frac{\eta_0}{m},
\EEQ
are the shear viscosity and thermal conductivity, respectively, in the elastic ($\alpha=1$) and smooth ($\beta=-1$) case.
Appendix \ref{appAB} shows that the $\gamma$ coefficients can be expressed in terms of collision integrals involving the HCS VDF $f^{(0)}$.

\subsection{Approximate form for $f^{(0)}$}

Thus far, we did not need in this section to specify the VDF $f^{(0)}$. Furthermore, the dependence on the number of degrees of freedom $\dt$ and $\dr$ in the equations above obeys to purely geometric considerations from the point of view that the explicit form of the collision rules has not been used yet. Now, as a second step in the quest for explicit expressions for the transport coefficients, we adopt the semi-Maxwellian approximation given by Eq.\ \eqref{3.1}, which implies that the cumulants $a_{20}^{(0)}$ and $a_{11}^{(0)}$ [see Eqs.\ \eqref{eq:66-67}] vanish. As a matter of fact, it has been previously observed \cite{SKS11,VSK14,VSK14b} that those cumulants are indeed generally small, at least in the HS case.
Preliminary results \cite{paperIII} show that the cumulants are also relatively small in the HD case, except for high inelasticity.

Equation \eqref{3.1} allows us to carry out the collision integrals in Eqs.\ \eqref{nueta_b} and \eqref{Y-Xi} by applying the collision rules, which include vector products (see Appendix \ref{appA}). This gives rise to a much subtler and complex dependence on the number of degrees of freedom $\dt$ and $\dr$ \cite{MS19b}, which we simplify under the constraints that the results remain being valid for three-dimensional rough HS ($\dt=\dr=3$), two-dimensional rough HD ($\dt=2$, $\dr=1$), and $d$-dimensional smooth particles ($\dt=d$, $\dr\to 0$). The algebra involved in the computation of the collision integrals is rather tedious, so here we only provide the final results. A summary of the main explicit expressions obtained by the combination of Eqs.\ \eqref{3.1} and \eqref{Sonine} is presented in Table \ref{table0}. Those expressions are equivalent, in the HS case  ($\dt=\dr=3$), to those shown in Table I of Ref.\ \cite{KSG14}.

{\renewcommand{\arraystretch}{2}
\begin{table}
\caption{Summary of the main explicit expressions in the approximations \eqref{3.1} and \eqref{Sonine}.}\label{table0}
\begin{ruledtabular}
\begin{tabular}{l}
$\displaystyle{\widetilde\a=\frac{1+\a}{2}}, \quad \displaystyle{\bt=\frac{1+\b}{2}\frac{\kappa}{1+\kappa}}$\\
$\displaystyle{\frac{\Tt^{(0)}}{T}=\tau_t=\frac{\dt+\dr}{\dt+\dr\theta}}, \quad \displaystyle{\frac{\Trot^{(0)}}{T}=\tau_r=\frac{\dt+\dr}{\dt/\theta+\dr}}$\\
$\displaystyle{\theta=\sqrt{\left[h-\frac{1}{2}\left(\frac{\dt}{\dr}-1\right)\right]^2+\frac{\dt}{\dr}}+h-\frac{1}{2}\left(\frac{\dt}{\dr}-1\right)}$\\
$\displaystyle{h\equiv \frac{\dt(1+\kappa)^2}{2\dr\kappa(1+\b)^2}\left[{1-\a^2}-\frac{1-\frac{\dr}{\dt}\kappa}{1+\kappa}(1-\b^2)\right]}$\\
$\displaystyle{\nu=K n\sigma^{\dt-1}\sqrt{2\tau_tT/m}}, \quad \displaystyle{K\equiv \frac{\sqrt{2}\pi^{\frac{\dt-1}{2}}}{\Gamma(\dt/2)}}$\\
$\displaystyle{\frac{\zeta^{(0)}}{\nu}=\zeta^*=\frac{1}{\dt+\dr\theta}\left[1-\a^2+\frac{\dr}{\dt}\frac{1-\b^2}{1+\kappa}
(\kappa+\theta)\right]}$\\[2mm]
\hline
 $\displaystyle{\eta=\frac{n\tau_t T}{\nu}\frac{1}{\nu_\eta^*-\frac{1}{2}\zeta^*}},\quad \displaystyle{\eta_b=\frac{\dr n\tau_t\tau_r T}{\dt\nu}\gamma_E}$\\
$\displaystyle{\lambda={\tau_t\lambda_t+\tau_r\lambda_r}}, \quad\displaystyle{\lambda_t=\frac{\dt+2}{2}\frac{n\tau_t T}{m\nu}\gamma_{A_t}}, \quad\displaystyle{\lambda_r=\frac{\dr}{2}\frac{n\tau_t T}{m\nu}\gamma_{A_r}}$\\
 $\displaystyle{\mu={\mu_t+\mu_r}}, \quad \displaystyle{\mu_t=\frac{\dt+2}{2}\frac{\tau_t^2 T^2}{m\nu}\gamma_{B_t}}, \quad \displaystyle{\mu_r=\frac{\dr}{2}\frac{\tau_t \tau_r T^2}{m\nu}\gamma_{B_r}}$\\
$\displaystyle{\xi=\frac{\dt\tau_t\xi_t+\dr\tau_r\xi_r}{\dt+\dr}=\gamma_E\Xi}, \quad \displaystyle{\xi_t=\gamma_E\Xi_t}, \quad \displaystyle{\xi_r=\gamma_E\Xi_r}$\\[1mm]
\hline
$\displaystyle{\nu_\eta^*=\frac{4}{\dt(\dt+2)}\Bigg[(\dt+3)\left(\at+\frac{\dr}{\dt}\bt\right)-3\at^2-\frac{d_\rot^2}{\dt}\bt^2}$\\
\hspace*{7mm}$\displaystyle{-\frac{4\dr\bt}{\dt-1}\left(\at-\frac{\bt\theta}{4\dt}\right)\Bigg]}$\\
$\displaystyle{\gamma_E=\frac{2}{\dt}\left({\Xi_t-\Xi_r-\frac{\dt+\dr}{2\dt}\zeta^*}\right)^{-1}}$\\
$\displaystyle{\Xi_t=\frac{3\dr\tau_r}{2d_\tr^2}\Bigg\{1-\a^2+\frac{\dr}{\dt}\frac{\kappa}{1+\kappa}(1-\b^2)}$\\ \hspace*{7mm}$\displaystyle{-\left(\frac{1+\b}{1+\kappa}\right)^2\frac{\kappa}{3}\left[\frac{\dr}{\dt}(\theta-3)-2\right]
\Bigg\}}$\\
$\displaystyle{\Xi_r=\frac{\tau_t}{2\dt}\frac{1+\b}{1+\kappa}\Bigg\{(1-\b)\left(\frac{\dr}{\dt}\theta-2\right)
}$\\
\hspace*{7mm}$\displaystyle{+\frac{1+\b}{1+\kappa}\kappa\left[\frac{\dr}{\dt}(\theta-3)-2\right]\Bigg\}}$\\
$\displaystyle{\Xi=\frac{3\dr\tau_t\tau_r}{2\dt(\dt+\dr)}\left\{1-\a^2+\frac{1-\b^2}{3(1+\kappa)}\left[\frac{\dr}{\dt}(3\kappa+\theta)-2\right]\right\}}$\\
$\displaystyle{\gamma_{A_t}=\frac{Z_r-Z_t-2\zeta^*}{\left(Y_t-2\zeta^*\right)\left(Z_r-2\zeta^*\right)-Y_r Z_t}}$\\
$\displaystyle{\gamma_{A_r}=\frac{Y_t-Y_r-2\zeta^*}{\left(Y_t-2\zeta^*\right)\left(Z_r-2\zeta^*\right)-Y_r Z_t}}$\\
$\displaystyle{\gamma_{B_t}=\zeta^*\frac{\gamma_{A_t}\left(Z_r-\frac{3}{2}\zeta^*\right)-\gamma_{A_r}Z_t}
{\left(Y_t-\frac{3}{2}\zeta^*\right)\left(Z_r-\frac{3}{2}\zeta^*\right)-Y_r Z_t}}$\\
$\displaystyle{\gamma_{B_r}=\zeta^*\frac{\gamma_{A_r}\left(Y_t-\frac{3}{2}\zeta^*\right)-\gamma_{A_t}Y_r}
{\left(Y_t-\frac{3}{2}\zeta^*\right)\left(Z_r-\frac{3}{2}\zeta^*\right)-Y_r Z_t}}$\\
$\displaystyle{Y_t=\frac{1}{\dt(\dt+2)}\left[(20+7\dt)\left(\at+\frac{\dr}{\dt}\bt\right)-3(\dt+8)\at^2\right.}$\\
\hspace*{7mm}$\displaystyle{\left.-\frac{(12+7\dt)\dr}{\dt}\bt^2-\frac{16\dr}{\dt}\at\bt-\bt^2\frac{\theta}{\kappa}\frac{\dr(\dt+4)}{\dt}\right]}$\\
$\displaystyle{Y_r=\frac{\dt+2}{d_\tr^2}\frac{\bt}{\kappa}\left(1-\frac{3\bt}{\theta}-\frac{\bt}{\kappa}\right)},\quad \displaystyle{Z_t=-\frac{2\dr}{d_\tr^2}\bt^2\frac{\theta}{\kappa}
}$\\
$\displaystyle{Z_r=\frac{2}{\dt}\left[\at+\frac{\dr}{\dt}\bt+\frac{\bt}{\kappa}\left(\frac{2\dt+1}{\dt}-\frac{\bt}{\kappa}-2{\bt}-\frac{4{\at}}{\dt}\right)\right]}$\\
\end{tabular}
\end{ruledtabular}
\end{table}
}

{\renewcommand{\arraystretch}{2}
\begin{table*}
\caption{Temperature ratios ($\theta$ and $\tau_t$), reduced cooling rate ($\zeta^*$), and reduced transport coefficients ($\eta^*$, $\eta_b^*$, $\lambda^*$, $\mu^*$, and $\xi$) in certain limits.}
\begin{ruledtabular}
\begin{tabular}{cccc}
Quantity& Purely smooth particles& Quasismooth limit & Perfectly rough and elastic particles:  \\
 & ($\dt=d$, $\dr\to 0$) & $(\eet\rightarrow -1)$ & Pidduck's limit $(\een=\eet=1)$ \\
\hline
$\theta$&Irrelevant&$\dfrac{\dt}{\dr}\dfrac{(1+\kappa)^2(1-\een^2)}{\kappa (1+\eet)^2}\to\infty$&$1$\\
$\tau_t$&$1$&$0$&$1$\\
$\zeta^*$&$\dfrac{1-\een^2}{d}$&$\dfrac{2(1+\eet)}{\dt(1+\kappa)}\to 0$&$0$\\
$\eta^{*}$     & $\displaystyle{\frac{8d}{(1+\een)[d(3+\een)+4(1-\een)]}}$  & $\displaystyle{\frac{4\dt(\dt-1)}{(1+\een)[2(d_\tr^2-1)+\dt(1-3\een)+2\een]}}$ & $\displaystyle{\frac{2\dt(1+\kappa)^2}{2\dt+(d_\tr^2+2\dt-2)\kappa}}$\\
$\eta_b^{*}$     & $0$ & $\displaystyle{\frac{8}{(\dt+2)(1-\een^2)}}$& $\displaystyle{\frac{2\dt\dr(1+\kappa)^2}{(\dt+2)(\dt+\dr)^2\kappa}}$\\
$\lambda^{*}$     & $\displaystyle{\frac{16(d-1)}{(1+\een)[d(3+5\een)-8\een]}}$ & $\displaystyle{\frac{4(\dt-1)(\dt+\dr)}{(\dt+2)^2(1+\een)}}$ & $\displaystyle{\frac{4\dr(1+\kappa)^2}{(\dt+2)^2}\frac{P_N(\kappa)}{P_D(\kappa)}}$\\
 $\mu^{*}$     & $\displaystyle{\frac{4(d+2)(1-\een)}{d(5+3\een)+4(1-3\een)}\lambda^{*}}$ & 0 & 0\\
$\xi$     & 0 & 0 & 0\\
\end{tabular}
\end{ruledtabular}
\label{table1}
\end{table*}
}

\begin{figure}
    \includegraphics[width=0.47\textwidth]{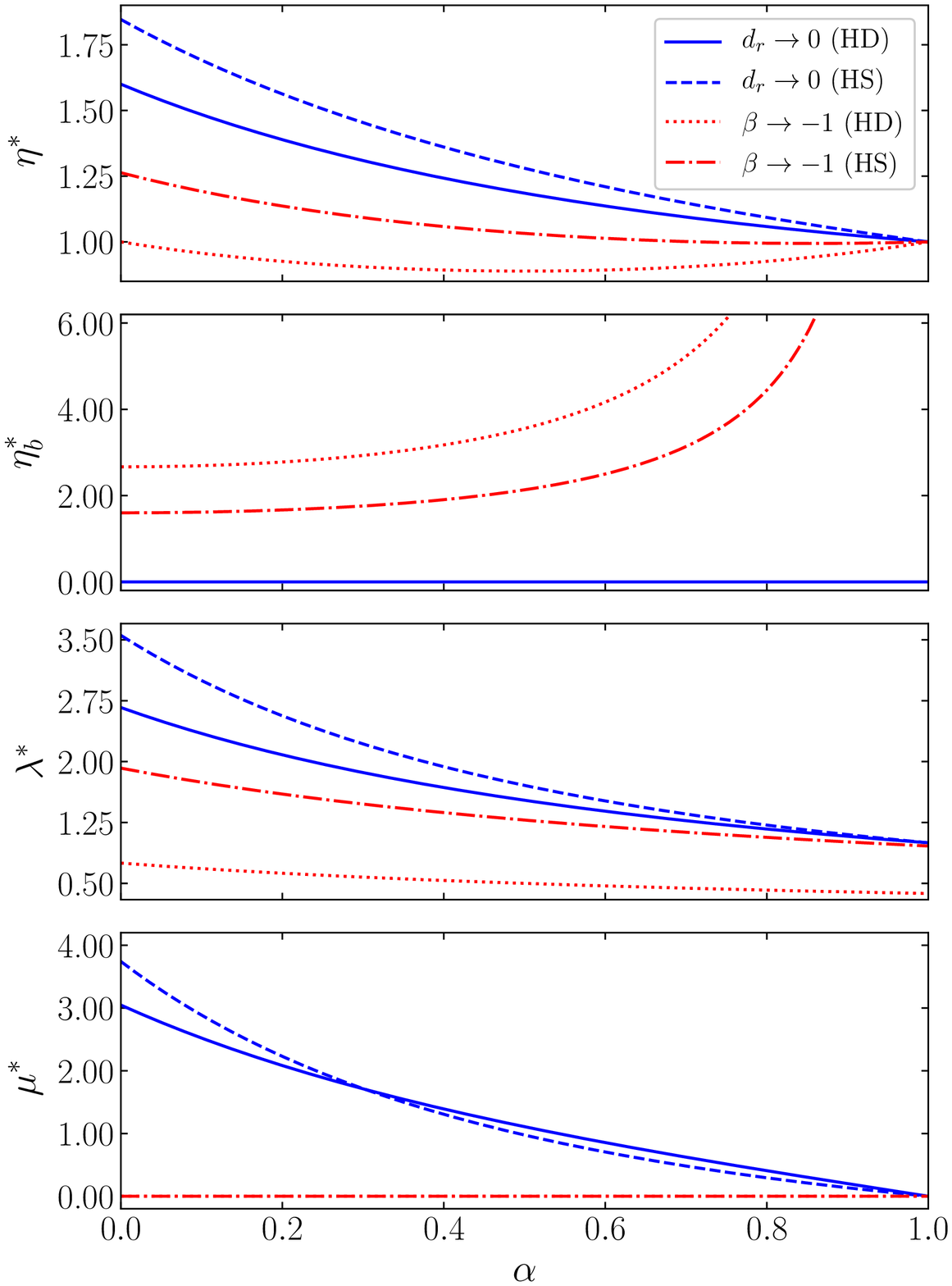}
    \caption{Dependence of the relevant (reduced) transport coefficients for the smooth ($\dr\to 0$) and quasismooth ($\beta\rightarrow -1$) limits on the coefficient of normal restitution for HS and HD granular gases. }
    \label{fig:multi_smooth}
\end{figure}

\begin{figure}
    \includegraphics[width=0.47\textwidth]{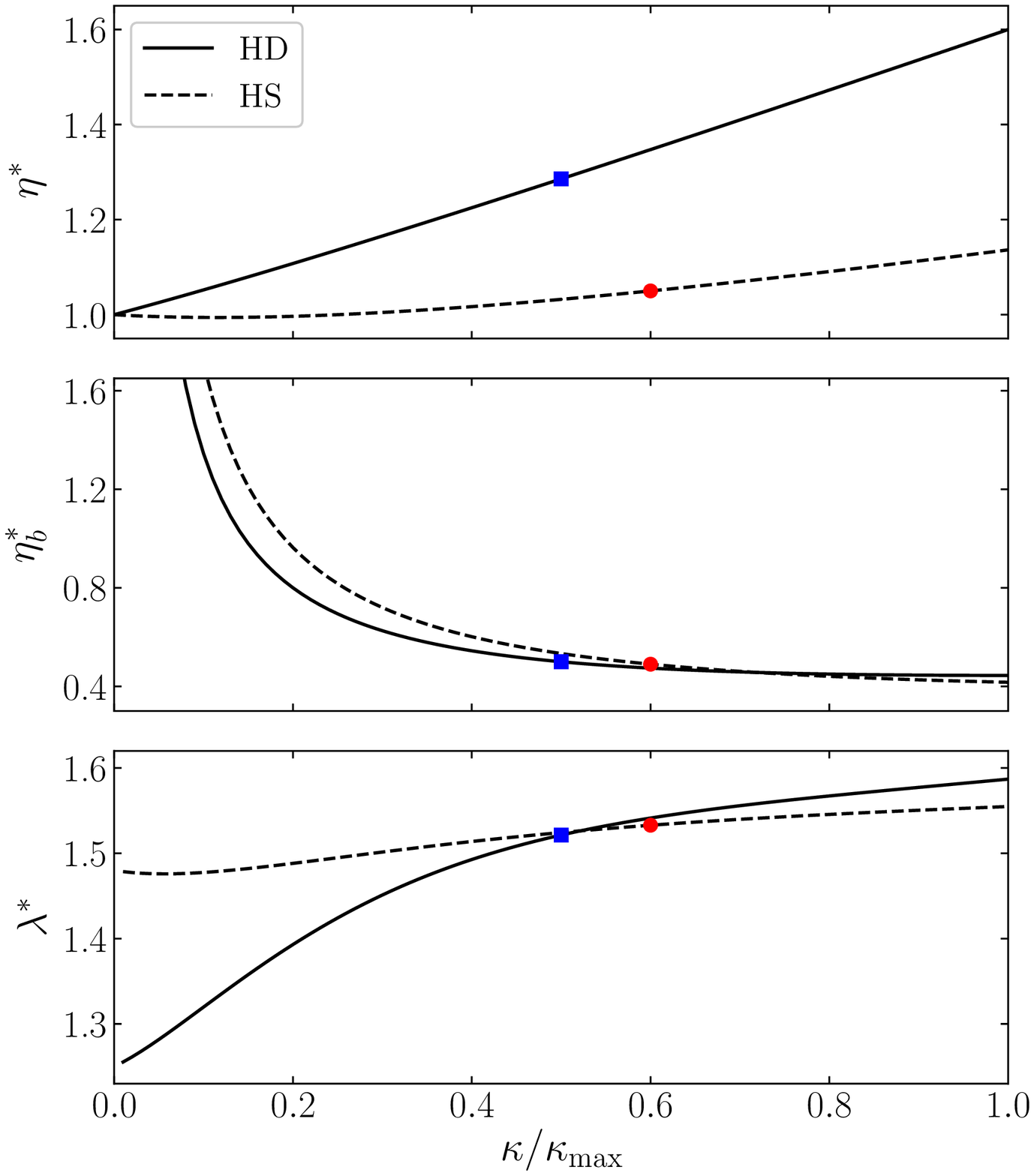}
    \caption{Dependence  of the nonzero (reduced) transport coefficients for  perfectly elastic and perfectly rough particles (Pidduck's gas) on the reduced moment of inertia $\kappa$, relative to its maximum value $\kappa_{\max}=1$ (HD) or $\kappa_{\max}=\frac{2}{3}$ (HS). The symbols correspond to uniform disks ($\kappa/\kappa_{\max}=\frac{1}{2}$)  and uniform spheres ($\kappa/\kappa_{\max}=\frac{3}{5}$). }
    \label{fig:multi_Pidduck}
\end{figure}
\section{Results}
\label{sec:5}

\subsection{Limiting cases}
While Table \ref{table0} gives the transport coefficients in terms of the coefficients of restitution ($\alpha$, $\beta$), the reduced moment of inertia ($\kappa$), and the number of degrees of freedom ($\dt$, $\dr$), it is interesting to consider some important limiting cases.

The first situation corresponds to a $d$-dimensional gas of \emph{smooth} particles. In that case, $\dt\to d$ and, given that  $\beta\to -1$ is a singular limit (see below), we formally take $\dr\to 0$.
Since the rotational-to-translational  temperature ratio lacks any physical meaning in the purely smooth case, its irrelevant precise value is not needed.
In fact,  on purely mathematical grounds, Eq.\ \eqref{theta} shows that $\lim_{\dr\to 0}\theta=\text{finite}$ if $\alpha>{|\beta-\kappa|}/{1+\kappa}$.
Upon taking the limit $\dr\to 0$ in Table \ref{table0}, one can easily obtain  the expressions shown in the second column of Table \ref{table1}. They agree with previous results \cite{BC01,GSM07} particularized to  the Maxwellian approximation. The same results are obtained by formally setting $\theta=0$ and  either  $\beta= -1$ or $\kappa= 0$, except that a spurious factor $\tau_t=1+\dr/\dt$ is attached to $\lambda$ and $\mu$ \cite{KSG14}.

As said before, the quasismooth limit $\beta\rightarrow -1$ is singular and completely different from the smooth case \cite{S11b,KSG14}. This distinction is physical and independent of the approximations carried out in this paper. The physical origin of the quasismooth singularity of the HCS can be summarized as follows. If the particles are strictly smooth ($\beta = -1$), then the rotational degrees of freedom
are quenched, so that the (physically irrelevant) rotational temperature remains constant while the translational temperature monotonically decreases with time.
The rotational-to-translational temperature ratio diverges but  there is no mechanism
transferring energy from the rotational to the translational degrees of freedom; in other words, the channel transferring energy between the rotational and translational degrees of freedom via collisions is broken if $\beta=-1$.  However, if $\beta=-1+\varepsilon$, where $0<\varepsilon\ll 1$, then the  rotational-to-translational temperature
ratio becomes so huge that it is eventually able to activate and ``feed'' the weak energy channel connecting the rotational and translational temperatures,  thus producing a nonnegligible effect on the HCS VDF \cite{SKG10,SKS11,VSK14,VSK14b}.

After carefully taking the limit $\beta\to -1$, the results for the quasismooth limit displayed  in the third column of Table \ref{table1} are obtained. As already noticed in Ref.\  \cite{KSG14}, $\theta\sim (1+\beta)^{-2}\to\infty$, $\zeta\sim (1+\beta)\to 0$, and no dependence on the reduced moment of inertia $\kappa$ remains in the transport coefficients after taking the quasismooth limit.

Figure \ref{fig:multi_smooth} shows the differences between the smooth and quasismooth (reduced) transport coefficients. In the cases of the shear viscosity $\eta^*$ and the thermal conductivity $\lambda^*$, we observe that those coefficients are higher for HS than for HD; additionally, they are higher for smooth particles (monotonic behavior) than in the quasismooth limit (nonmonotonic behavior).
In what respects the bulk shear viscosity $\eta_b^*$, it vanishes for smooth particles, but not in the quasismooth limit, in which case it takes higher values for HD than for HS.
Finally, the Dufour-like coefficient $\mu^*$ vanishes in the quasismooth limit, but not for smooth particles, the HD value being larger than the HS one if $\alpha>0.303$.

As a third limiting situation, we now consider  a system of particles perfectly elastic ($\alpha=1$) and perfectly rough ($\beta=1$). Since energy is conserved by collisions [see Eq.\ \eqref{eq:7}], the equipartition principle holds. In the HS case, this system was first introduced about one hundred years ago by Pidduck \cite{P22} and is frequently used to model polyatomic molecules \cite{CC70,MSD66,K10}. The results for HS and HD gases are given in the fourth column of Table \ref{table1}. In the case of $\lambda$,   $P_N(\kappa)=N_0+N_1\kappa+N_2\kappa^2$ and $P_D(\kappa)=D_0+D_1\kappa+D_2\kappa^2+D_3 \kappa^3$ are polynomials with coefficients $(N_0,N_1,N_2,D_0,D_1,D_2,D_3)=(10,39,24,2,11,12,21)$ and $(37,151,50,12,75,101,102)$ for HD and HS, respectively.
It must be noted that, when setting $\alpha=\beta=1$ in the expressions of Table \ref{table0}, we took the licence of using $\dr=\frac{1}{2}\dt(\dt-1)$   to simplify the final results for $\eta^*$ and $\lambda^*$. Actually, the relation $\dr=\frac{1}{2}\dt(\dt-1)$
is exact due to the relation of rotational mechanics on a $\dt$-translational geometry and the orthogonal group $O(\dt)$ \cite{H15,GM21}.

The dependence of $\eta^*$, $\eta_b^*$, and $\lambda^*$ on the reduced moment of inertia for the HS and HD Pidduck gases is displayed in Fig.\  \ref{fig:multi_Pidduck}. Given a common value of $\kappa/\kappa_{\max}$, while the shear viscosity is higher for HD than for HS, the opposite happens in the case of the bulk viscosity (except if $\kappa/\kappa_{\max}>0.718$, in which case the HD curve is slightly above the HS one). The thermal conductivity is higher for HD than for HS only if $\kappa/\kappa_{\max}>0.522$. If the particles have a uniform mass distribution, then the HD-to-HS ratios are equal to $1.22$, $1.02$, and $0.99$ for $\eta^*$, $\eta_b^*$, and $\lambda^*$, respectively.

\begin{figure}
    \includegraphics[width=\columnwidth]{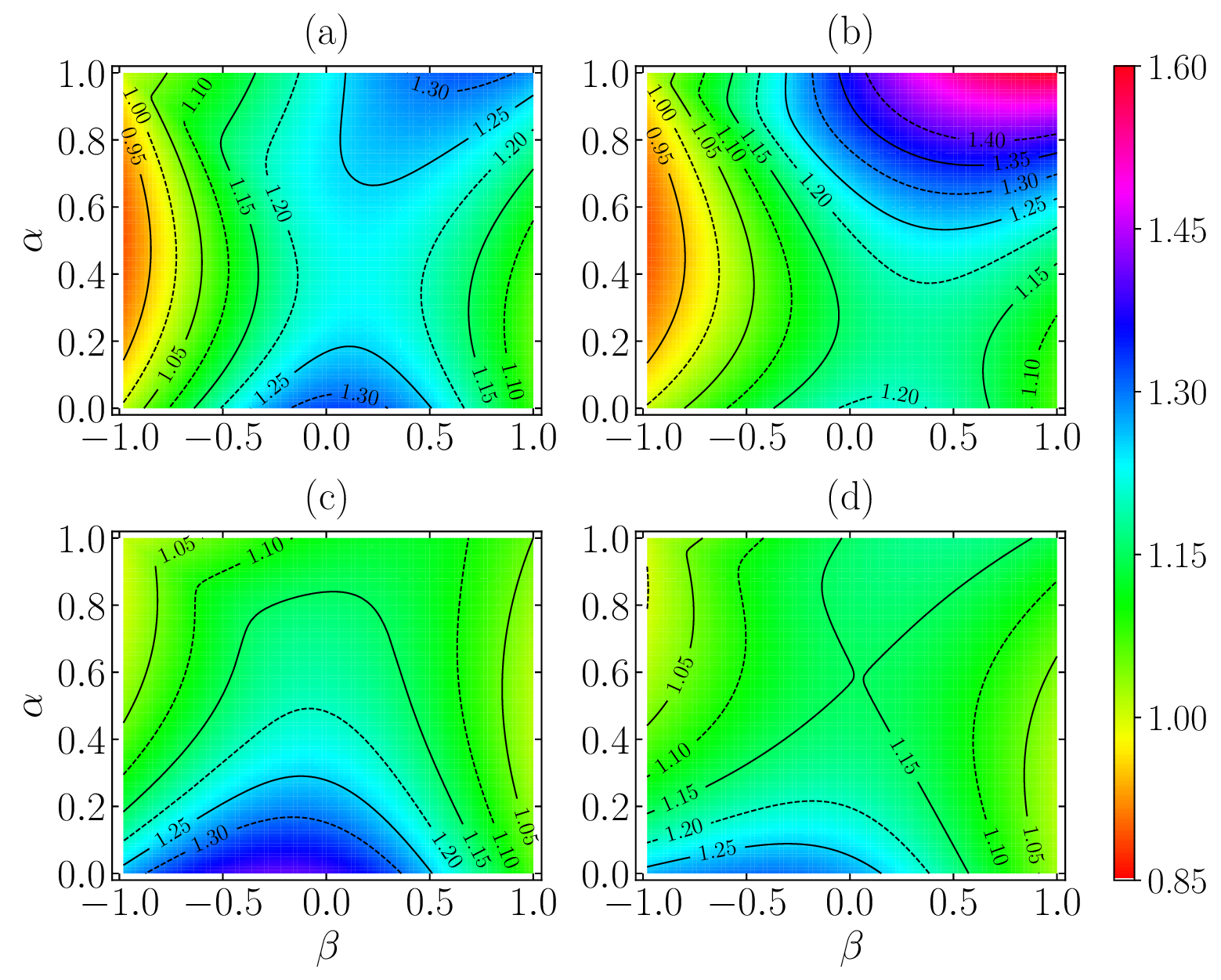}
    \caption{Density plots of the reduced shear viscosity $\eta^*$ in the plane $\beta$ vs $\alpha$ for (a) HD with a uniform mass distribution ($\kappa=\frac{1}{2}$), (b) HD with a  mass distribution concentrated on the outer surface ($\kappa=1$), (c) HS with a uniform mass distribution ($\kappa=\frac{2}{5}$), and (d) HS with a  mass distribution concentrated on the outer surface ($\kappa=\frac{2}{3}$). }
    \label{fig:eta}
\end{figure}

\begin{figure}
    \includegraphics[width=\columnwidth]{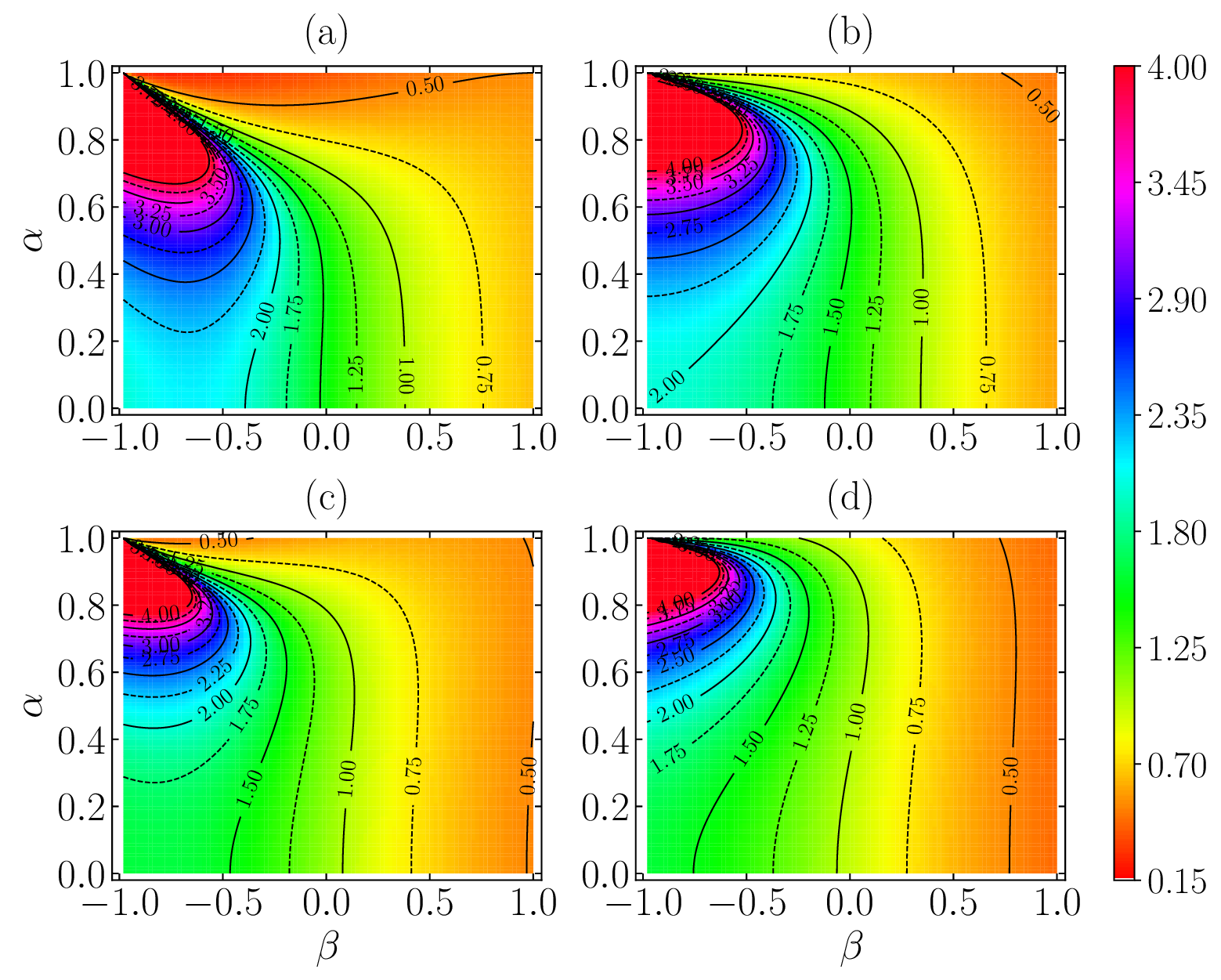}
    \caption{Same as described in the caption of Fig.\ \ref{fig:eta} but for the reduced bulk viscosity $\eta_b^*$. }
    \label{fig:etab}
\end{figure}

\begin{figure}
    \includegraphics[width=\columnwidth]{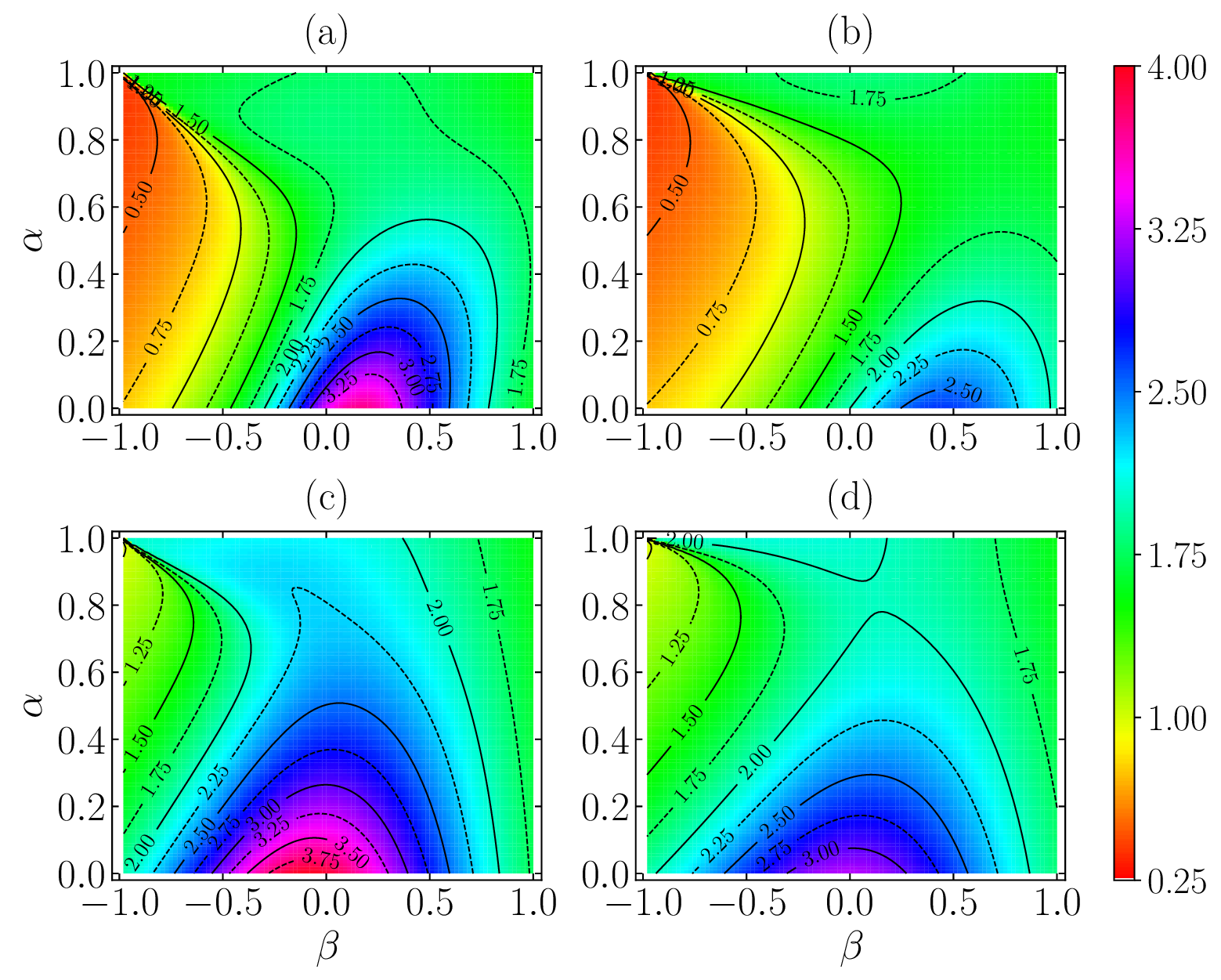}
    \caption{Same as described in the caption of  Fig.\ \ref{fig:eta} but for the thermal conductivity $\lambda^*$. }
    \label{fig:lambda}
\end{figure}

\begin{figure}
    \includegraphics[width=\columnwidth]{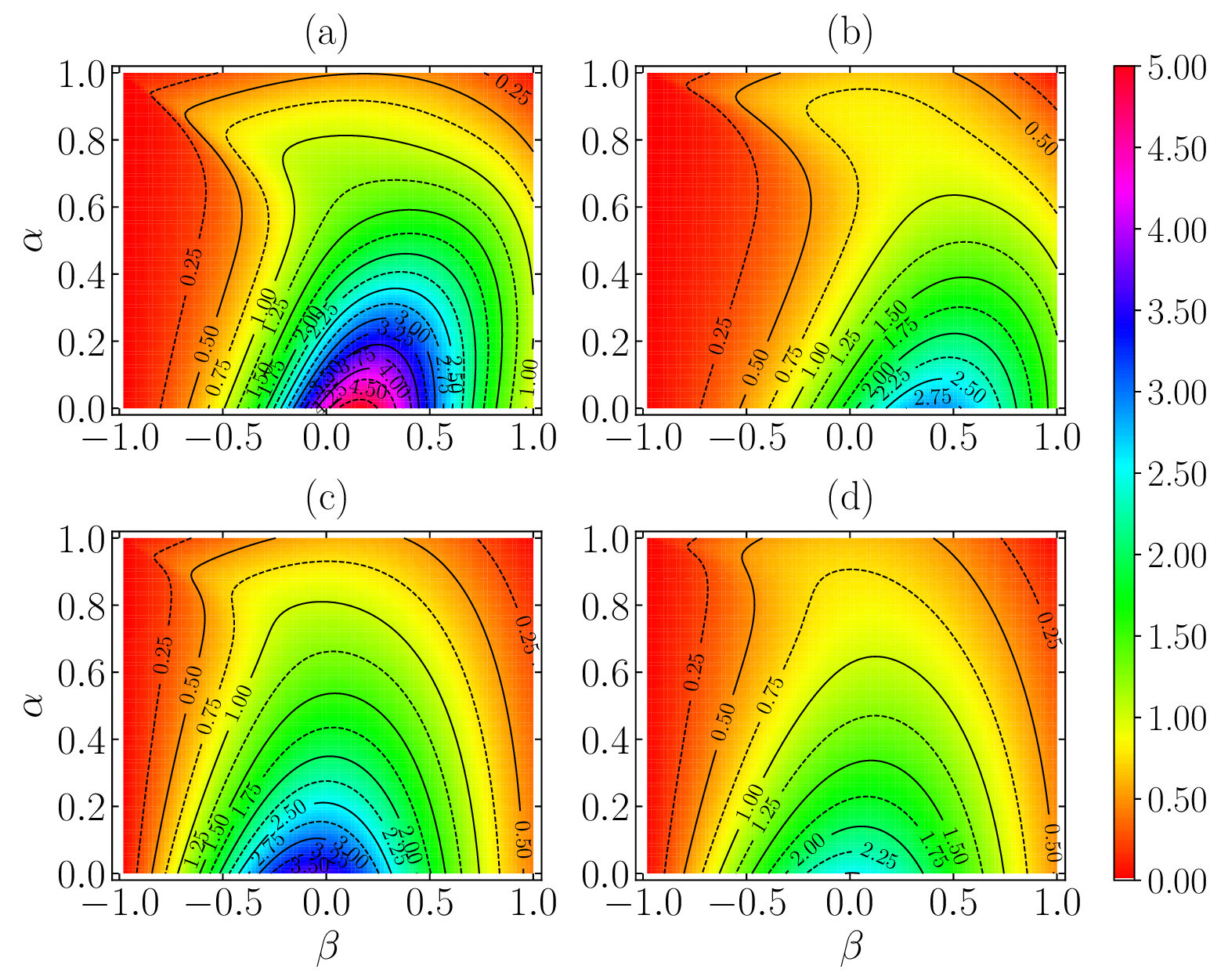}
    \caption{Same as described in the caption of  Fig.\ \ref{fig:eta} but for the Dufour-like coefficient $\mu^*$. }
    \label{fig:mu}
\end{figure}

\begin{figure}
    \includegraphics[width=\columnwidth]{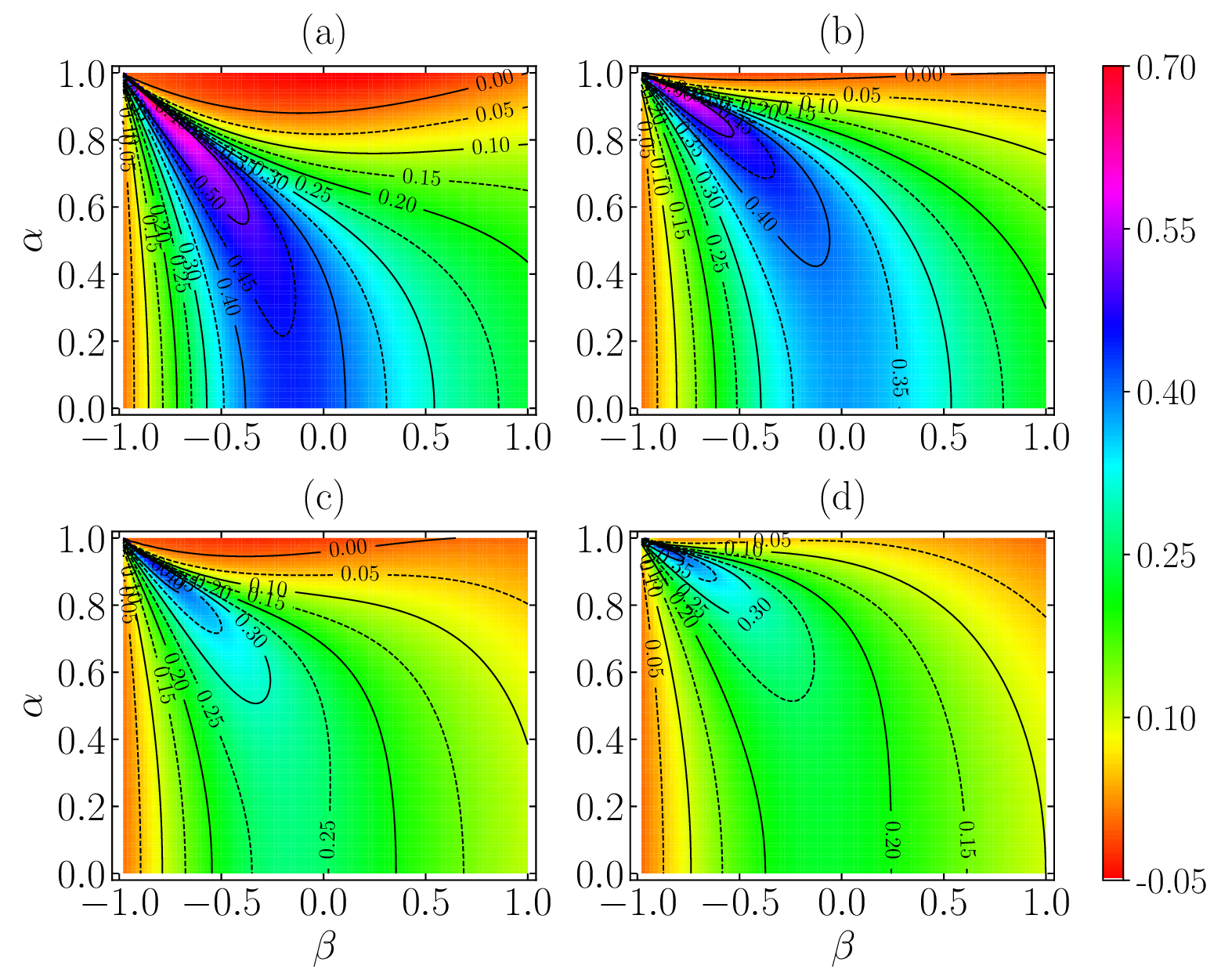}
    \caption{Same as described in the caption of  Fig.\ \ref{fig:eta} but for the velocity-divergence transport coefficient $\xi$. }
    \label{fig:xi}
\end{figure}

\subsection{General system}
Now we go back to the general case and illustrate the dependence of the five transport coefficients $\eta^*$, $\eta_b^*$, $\lambda^*$, $\mu^*$, and $\xi$ on the coefficients of restitution ($\alpha$, $\beta$) and the reduced moment of inertia ($\kappa$) for both HS and HD granular gases. The results are displayed as density plots in Figs.\ \ref{fig:eta}--\ref{fig:xi}. Two characteristic cases of mass distribution are considered: uniform distribution ($\kappa=\frac{1}{2}$ and $\frac{2}{5}$ for HD and HS, respectively) and mass concentrated on the outer surface ($\kappa=\kappa_{\max}=1$ and $\frac{2}{3}$ for HD and HS, respectively).

We observe an intricate influence of both $\alpha$ and $\beta$ on the transport coefficients, with typically a strong nonmonotonic dependence on $\beta$ with, at fixed $\alpha$, a single maximum around $\beta\approx 0$ for $\eta^{*}$, $\lambda^{*}$, and $\mu^{*}$, a maximum near $\beta\to -1$ for $\eta_b^{*}$, and a more complex behavior for $\xi$. Note that the bulk viscosity reaches very high values in the quasielastic and quasismooth region (see Fig.\ \ref{fig:etab}); in fact, as shown in Fig.\ \ref{fig:multi_smooth} and Table \ref{table1}, $\eta_b^*$ diverges in the combined limit $\beta\to -1$ and $\alpha\to 1$.
Moreover, $\eta^*$ and $\lambda^*$ are the transport coefficients more sensitive to the dimensionality and to the mass distribution.
It is also worth mentioning that $\xi$ reaches negative values in a narrow lobe region near $\alpha=1$ (see Fig.\ \ref{fig:xi}). That region is wider for HD than for HS and tends to shrink as the moment of inertia grows; in fact, it disappears  for HS with a surface mass distribution.

\section{Concluding remarks}\label{sec:6}

In this study, we have considered a model of a granular gas as composed by spherical particles with constant coefficients of normal ($\een$) and tangential ($\eet$) restitution. Previous results \cite{KSG14} for the transport coefficients  of a dilute gas of inelastic and rough HS have been complemented  with novel results for the parallel case of HD. We have developed this analysis in a unified vector space framework, based on previous works \cite{MS19,MS19b}, which allows us to obtain general expressions in terms of the number of translational ($\dt$) and rotational ($\dr$) degrees of freedom. The choice of the embedding Euclidean three-dimensional space is essential to get results for both geometries within a common framework. Particular aspects, especially the orthogonality condition between rotational and translational velocities in the HD case, permit us to neglect the computation of certain quantities and reduce them to already known HS terms being parameterized by $\dt-2$ or $(\dr-1)/2$ prefactors, as convenience, in the spirit of Refs.\ \cite{MS19,MS19b}.

The analysis has been carried out in the context of the nonlinear Boltzmann equation, where the system is assumed to be fully described by the one-particle VDF $f(\rr,\vv,\oo;t)$. Under the assumptions of (i) small gradients of the hydrodynamic fields  [number density $n(\rr,t)$, flow velocity $\uu(\rr,t)$, and  granular temperature $T(\rr,t)$] and (ii) a ``normal'' solution (i.e., the space and time dependence of the one-particle VDF takes place through a functional dependence on  $n$, $\uu$, and $T$), the Chapman--Enskog method has been used to solve the  Boltzmann equation up to first order in the gradients. In that way, the NSF hydrodynamic equations are obtained by supplementing the balance equations for mass [Eq.\ \eqref{eq:17}], momentum [Eq.\ \eqref{eq:18}], and energy  [Eq.\ \eqref{eq:23}] with constitutive equations for the pressure tensor $P_{ij}(\rr,t)$ [Eq.\ \eqref{eq:52}], the heat flux $\qheat(\rr,t)$ [Eq.\ \eqref{eq:53}], and the cooling rate $\zeta(\rr,t)$ [Eq.\ \eqref{eq:45c}].

The derivation of the  associated transport coefficients has been carried out along three successive stages of increasing concreteness and level of approximation. In a first stage, the transport coefficients are expressed [see Eqs.\ \eqref{eq:73-65} and \eqref{eq:69-72}] in terms of collision integrals involving the zeroth-order (HCS) VDF $f^{(0)}$ and the functions $\Ab$, $\Bb$, $\Cbij$, and $\Eb$ characterizing the first-order VDF $f^{(1)}$ [see Eq.\ \eqref{eq:44}]; those functions are the solutions of a set of linear integral equations [see Eqs.\ \eqref{eq:48a-d}] with inhomogeneous terms related to $f^{(0)}$ [see Eqs.\  \eqref{eq:42a-d}]. Next, in a second stage, Sonine-like forms for the functions  $\Ab$, $\Bb$, $\Cbij$, and $\Eb$ are assumed [see Eqs.\ \eqref{Sonine}], with coefficients ($\gamma_{A_t}$, $\gamma_{A_r}$, $\gamma_{B_t}$, $\gamma_{B_r}$, $\gamma_{C}$, and $\gamma_{E}$) that can be expressed in terms of collisional integrals involving $f^{(0)}$ [see Eqs.\ \eqref{Y-Xi} and \eqref{gammas}]. Finally, as  the third and final stage, the HCS VDF $f^{(0)}$ is approximated by the product of a Maxwellian translational VDF times the marginal rotational VDF [see Eq.\ \eqref{3.1}]. The resulting explicit expressions for the transport coefficients as functions of the coefficients of restitution ($\alpha$ and $\beta$), the reduced moment of inertia ($\kappa$), and the numbers of degrees of freedom ($\dt$ and $\dr$) are displayed in Table \ref{table0}. In general, the transport coefficients exhibit a rather complex nonlinear dependence on $\een$, $\eet$,  and $\kappa$, as  exposed in the density plots of Figs.\ \ref{fig:eta}--\ref{fig:xi}.

The choice $(\dt,\dr)=(3,3)$ allows us to recover known results for three-dimensional HS \cite{KSG14}, except that in our approach we did not need to assume a Maxwellian form for the marginal rotational VDF. Moreover, novel results for two-dimensional HD are derived via the choice $(\dt,\dr)=(2,1)$. Thus, the outcome quantities can be used as a unified set of formulas for  theoretical and experimental researchers, as well as  a source of comparison between HD and HS setups.

Some special limiting cases have been exposed in Table\ \ref{table1}: smooth, quasismooth, and Pidduck's  limits. The common description in terms of translational and rotational degrees of freedom let a direct recovery of the purely smooth case results by formally taking the limit $\dr\to 0$ at fixed $\beta$, thus circumventing  the singular nature of the quasismooth limit $\beta\to -1$. In the latter limit, a universal lack of dependence on $\kappa$ of the transport coefficients, already seen for HS \cite{KSG14}, is observed. The quasismooth limit is quite distinct from the purely smooth case, as  shown in Fig.\ \ref{fig:multi_smooth}.  Furthermore, we have extended the original Pidduck's system \cite{P22} ($\een=\eet=1$) to our description, and novel results for HD are obtained;  the dependencies with the reduced moment of inertia are shown in Fig. \ref{fig:multi_Pidduck}, where one can observe that the values of the coefficients $\eta_b^*$ and $\lambda^*$ with a uniform mass distribution are similar for the two considered setups. It is also interesting to remark that the transport coefficient ($\xi$) associated with the velocity-divergence correction of the cooling rate vanishes in all these limits, for both HD and HS, as expected.

An immediate application of this work is the use of the closed set of NSF hydrodynamic equations to analyze the stability of the HCS, again in a unified framework encompassing the special HS and HD cases. This is the subject of the companion paper \cite{paperII}. Additionally, the extension of the results to stochastically driven granular gases is straightforward (since the evaluation of the collision integrals has already been done in the present paper) and will be published elsewhere. Another future goal of our research is to go back to the second stage mentioned above and assume a form for $f^{(0)}$ where excess velocity kurtoses and translational-rotational velocity correlations are not neglected. Preliminary results \cite{paperIII} are quite promising.

Lastly, we hope that this research will inspire future works in the field, which could provide simulation and experimental results to compare with, as well as the introduction of alternative collisional models to describe systems of inelastic and rough particles.

\begin{acknowledgments}
The authors acknowledge financial support from the Grant No.\ PID2020-112936GB-I00/AEI/10.13039/501100011033 and from the Junta de Extremadura (Spain) through Grants No.\ IB20079 and No.\ GR18079, all of them partially financed by Fondo Europeo de Desarrollo Regional funds. A.M. is grateful to the Spanish Ministerio de Ciencia, Innovaci\'on y Universidades for support from a predoctoral fellowship Grant No.\ FPU2018-3503.
\end{acknowledgments}

\appendix
\section{Collision rules}
\label{appA}
The direct binary collision rules read
\BEQ\label{eq:1}
    m\vv_{1,2}^\prime=m\vv_{1,2}\mp \QQ, \quad
    I\oo_{1,2}^\prime=I\oo_{1,2}-\frac{\sigma}{2}\s\times\QQ,
\EEQ
where  $\QQ$ is the impulse that particle $1$ exerts on particle $2$. Our collision model is based on the existence of two constant coefficients of restitution, normal ($0<\alpha\leq 1$) and tangential ($-1\leq\beta\leq 1$), which are defined by the following relations:
\BEQ\label{eq:2}
    \s\cdot\grel^\prime = -\een (\s\cdot\grel), \quad \s\times\grel^\prime = -\eet (\s\times\grel).
\EEQ
Here,
\BEQ\label{eq:3}
    \grel = \vvab - \s\times \Ssab
\EEQ
is the relative velocity of the contact points at the moment of collision,  $\vvab\equiv \vva-\vvb$ being the center-of-mass  relative velocity and $\Ssab = \sigma(\ooa+\oob)/2$ being directly related to  the center-of-mass angular velocity.
Then, from the conservation of angular and linear momenta in each collision,  the impulse can be expressed as \cite{KSG14}
\BEQ\label{eq:4}
    \QQ = m\en(\s\cdot\vvab)\s-m\et\s\times(\s\times\vvab+\Ssab),
\EEQ
where
\BEQ\label{eq:5}
    \en \equiv\frac{1+\alpha}{2}, \qquad \et \equiv \frac{\kappa}{1+\kappa}\frac{1+\beta}{2}.
\EEQ

The loss of energy due to inelasticity and roughness is observed in the change of total kinetic energy, which is given by
\bal \label{eq:7}
    \Deltacoll E_K \equiv & \frac{m}{2}\Deltacoll(\va^2+\vb^2) + \frac{I}{2}\Deltacoll(\oa^2+\ob^2)\nn
    =& -\frac{m}{4}(1-\eet^2) \frac{\kappa}{1+\kappa}\left[\s\times(\s\times\vvab+\Ssab)\right]^2\nn &-\frac{m}{4}(1-\alpha^2)(\s\cdot\vvab)^2,
\eal
where $\Deltacoll \psi(\vv,\oo) \equiv \psi(\vv',\oo')-\psi(\vv,\oo)$. One can observe that, except if $\een = 1$ and either $\eet = -1$ or $\eet = 1$, the total kinetic energy is dissipated upon collisions. This expected fact is translated into a decay of the total granular temperature in Sec.\ \ref{subsec:2.2}.

The previous equations apply to both HS and HD. In the HS case, the translational velocity $\vv=v_x\widehat{\mathbf{i}}+v_y\widehat{\mathbf{j}}+v_z\widehat{\mathbf{k}}$ and the angular velocity  $\oo=\omega_x\widehat{\mathbf{i}}+\omega_y\widehat{\mathbf{j}}+\omega_z\widehat{\mathbf{k}}$ have $\dt=3$ and $\dr=3$ nontrivial components, respectively. However, in the HD case, $\vv=v_x\widehat{\mathbf{i}}+v_y\widehat{\mathbf{j}}$ and   $\oo=\omega_z\widehat{\mathbf{k}}$ have $\dt=2$ and $\dr=1$ nontrivial components, respectively, what simplifies the collision rules \cite{S18}. An important consequence of the distinction between spheres and disks is that the Jacobian of the transformation between pre- and postcollisional velocities turns out to depend on $\dr$, namely
\beq
\label{Jacob}
\left|\frac{\partial (\vva^\prime,\vvb^\prime,\ooa^\prime,\oob^\prime)}{\partial (\vva,\vvb,\ooa,\oob)}\right|=\alpha |\beta|^{2\dr/\dt}.
\eeq

\section{The $\gamma$ coefficients in terms of collision integrals}
\label{appAB}

When  Eqs.\ \eqref{Sonine} are inserted into Eqs.\ \eqref{eq:69-72} and \eqref{eq:45b}, one obtains
\begin{subequations}
\label{nueta_b}
\beq
\label{nueta_ba}
\nueta^*\equiv \frac{\nueta}{\nu}=\frac{\Lambda[V_iV_j-\frac{1}{\dt}V^2\delta_{ij}|\left(\frac{1}{\dt}\delta_{ij}\cc\cdot\partial_\cc-c_j\partial_{c_i}\right)f^{(0)}]}{\frac{1}{2}(\dt-1)(\dt+2) n\vth^2},
\eeq
\beq
    \nultr^{*}\equiv \frac{\nultr}{\nu} = Y_\tr + \frac{\gamma_{A_\rot}}{\gamma_{A_\tr}} Z_\tr,
    \eeq
\beq
    \nulrot^{*}\equiv \frac{\nulrot}{\nu} = \frac{{\gamma_{A_\tr}}Y_\rot + {\gamma_{A_\rot}}Z_\rot}{\gamma_{A_\rot}+\gamma_{A_\tr}\frac{a_{11}^{(0)}}{1+a_{11}^{(0)}}},
\eeq
\beq
  \numtr^{*}\equiv \frac{\numtr}{\nu} = Y_\tr + \frac{\gamma_{B_\rot}}{\gamma_{B_\tr}} Z_\tr,
\eeq
\beq
    \numrot^{*}\equiv \frac{\numrot}{\nu} = \frac{{\gamma_{B_\tr}}Y_\rot + {\gamma_{B_\rot}}Z_\rot}{\gamma_{B_\rot}+\gamma_{B_\tr}\frac{a_{11}^{(0)}}{1+a_{11}^{(0)}}},
\eeq
\BEQ
\label{Xi}
    \xi_\tr = \gamma_E \Xi_\tr , \quad \xi_\rot = \gamma_E \Xi_\rot,
\EEQ
\end{subequations}
where
\begin{subequations}
\label{Y-Xi}
\beq
Y_\tr=\frac{\Lambda[V^2V_i|\left(\partial_{c_i}-c_i\partial_\cc\cdot\cc\right)f^{(0)}]}{\frac{1}{2}(\dt+2)\nu n\vth^3\left[1+2a_{20}^{(0)}\right]},
\eeq
\beq
Z_\tr=-\frac{\Lambda[V^2V_i|c_i\partial_{\ww}\cdot\ww]}{\frac{1}{2}(\dt+2)\nu n\vth^3\left[1+2a_{20}^{(0)}\right]},
\eeq
\beq
Y_\rot=\frac{\Lambda[\omega^2V_i|\left(\partial_{c_i}-c_i\partial_\cc\cdot\cc\right)f^{(0)}]}{\frac{1}{2}\dt\dr\nu n\vth\oth^2\left[1+a_{11}^{(0)}\right]},
\eeq
\beq
Z_\rot=-\frac{\Lambda[\omega^2V_i|c_i\partial_{\ww}\cdot\ww]}{\frac{1}{2}\dt\dr\nu n\vth\oth^2\left[1+a_{11}^{(0)}\right]},
\eeq
\beq
\Xi_\tr=\dr\tau_\tr\tau_\rot\frac{\Lambda[V^2|\left(\frac{\partial_{\ww}\cdot\ww}{\dr\tau_\rot}-\frac{\dt+\cc\cdot\partial_\cc}{\dt\tau_\tr}\right)f^{(0)}]}{\dt\nu n\vth^2},
\eeq
\beq
\Xi_\rot=\dr\tau_\tr\tau_\rot\frac{\Lambda[\omega^2|\left(\frac{\partial_{\ww}\cdot\ww}{\dr\tau_\rot}-\frac{\dt+\cc\cdot\partial_\cc}{\dt\tau_\tr}\right)f^{(0)}]}{\dt\nu n\oth^2}.
\eeq
\end{subequations}
The six quantities in Eqs.\ \eqref{Y-Xi}, together with $\nueta^*$ in Eq.\ \eqref{nueta_ba}, define the fundamental collision integrals within the approximation given by Eqs.\ \eqref{Sonine}.

By comparing Eqs.\ \eqref{4.2a-e} to the second equalities in Eqs.\ \eqref{eq:73-65}, one obtains, after some algebra,
\begin{subequations}
\label{gammas}
\beq
    \gamma_C = \frac{2}{\nueta^{*}-\frac{1}{2}\zeta^{*}}, \quad \gamma_E = \frac{2/\dt}{\Xi_\tr-\Xi_\rot-\frac{\dt+\dr}{2\dt}\zeta^{*}},
\eeq
\beq
\gamma_{A_t} = \frac{{\Zr}-{\Zt}(1+\widetilde{a}_{11})-2\zeta^{*}}{\left({\Yt}-2\zeta^{*}\right)\left({\Zr}-2\zeta^{*}\right)-\left(\Yr-2\zeta^{*}\widetilde{a}_{11}\right){\Zt}},
\eeq
\beq
\gamma_{A_r} = \frac{{\Yt}(1+\widetilde{a}_{11})-{\Yr}-2\zeta^{*}}{\left({\Yt}-2\zeta^{*}\right)\left({\Zr}-2\zeta^{*}\right)-\left(\Yr-2\zeta^{*}\widetilde{a}_{11}\right){\Zt}},
\eeq
\begin{widetext}
\beq
\gamma_{B_t} = \frac{\zeta^{*}
\left[\gamma_{A_t}\left(
{\Zr}-\frac{3}{2}\zeta^{*}-{\Zt}\widetilde{a}_{11}
\right)-\gamma_{A_r}{\Zt} \right]
+\left({\Zr}-\frac{3}{2}\zeta^{*}\right)\widetilde{a}_{20}- {\Zt}\widetilde{a}_{11}
}{\left({\Yt}-\frac{3}{2}\zeta^{*}\right)\left({\Zr}-\frac{3}{2}\zeta^{*}\right)-\left(\Yr-\frac{3}{2}\zeta^{*}\widetilde{a}_{11}\right){\Zt}},
\eeq
\beq
    \gamma_{B_r} = \frac{\zeta^{*}\left[\gamma_{A_r}\left({\Yt}-\frac{3}{2}\zeta^{*}\right)
    -\gamma_{A_t}\left({\Yr}-{\Yt}\widetilde{a}_{11}\right)
    \right]
    +\left[{\Yt}-\frac{3}{2}\zeta^{*}(1-\widetilde{a}_{20})\right]\widetilde{a}_{11}-{\Yr}\widetilde{a}_{20}
    }
  {\left({\Yt}-\frac{3}{2}\zeta^{*}\right)\left({\Zr}-\frac{3}{2}\zeta^{*}\right)-\left(\Yr-\frac{3}{2}\zeta^{*}\widetilde{a}_{11}\right){\Zt}},
\eeq
\end{widetext}
\end{subequations}
where
\beq
\widetilde{a}_{11}=\frac{a_{11}^{(0)}}{1+a_{11}^{(0)}},\quad \widetilde{a}_{20}=\frac{a_{20}^{(0)}}{1+2a_{20}^{(0)}}.
\eeq


%

\end{document}